\documentclass[preprint]{elsarticle}

\usepackage[margin=3cm]{geometry}
\usepackage{booktabs}
\usepackage{amsmath}
\usepackage{amssymb}
\usepackage{algpseudocode}
\usepackage{arydshln}
\usepackage{algorithm2e}
\usepackage{caption}
\usepackage{graphicx}
\usepackage{epstopdf}
\usepackage{color,soul}
\usepackage{bm}
\usepackage{xcolor}
\usepackage{multirow, makecell, caption}
\usepackage{mathrsfs}
\usepackage[acronym]{glossaries}
\usepackage{optidef}
\usepackage{appendix}

\newglossaryentry{entryOne}
{
    name=Glossary Entry,
    description={Glossary entries are used to provide definitions for words in your document}
}

\begin{document}
\begin{frontmatter}

\title{A Graph Neural Network Surrogate Model for Multi-Objective Fluid-Acoustic Shape Optimization}

\author[label1]{Farnoosh Hadizadeh}
\ead{farnoosh.hadizadeh@ubc.ca}

\author[label2]{Wrik Mallik}
\ead{wrik.mallik@glasgow.ac.uk}

\author[label1]{Rajeev K. Jaiman}

\address[label1]{Department of Mechanical Engineering, The University of British Columbia, Vancouver, BC V6T 1Z4, Canada}
\address[label2]{James Watt School of Engineering, The University of Glasgow, Glasgow G12 8QQ, United Kingdom}
\ead{rjaiman@mech.ubc.ca}

\begin{abstract}

 This article presents a graph neural network (GNN) based surrogate modeling approach for fluid-acoustic shape optimization. The GNN model transforms mesh-based simulations into a computational graph, enabling global prediction of pressure and velocity flow fields around solid boundaries. We employ signed distance functions to implicitly represent geometries on unstructured nodes represented by the graph neural network. The trained graph neural network is employed here to predict the flow field around various airfoil shapes. The median relative error in the prediction of pressure and velocity for 300 test cases is 1-2\%. The predicted flow field is employed to extract the fluid force coefficients and the velocity profile of the boundary layer. The boundary layer velocity profile is then used to predict the flow field and noise levels, allowing the direct integration of the coupled fluid-acoustic analysis in the shape optimization algorithm. The fluid-acoustic shape optimization is extended to multi-objective shape optimization by minimizing trailing edge noise while maximizing the aerodynamic performance of airfoil surfaces. The results show that the overall sound pressure level of the optimized airfoil decreases by 13.9\% (15.82 dBA), and the lift coefficient increases by 7.2\%, for a fixed set of operating conditions. The proposed GNN-based integrated surrogate modeling with the shape optimization algorithm exhibits a computational speed-up of three orders of magnitude compared to while maintaining reasonable accuracy compared to full-order online optimization applications. The GNN-based surrogate model offers an efficient computational framework for fluid-acoustic shape optimization via adaptive morphing of structures.
\end{abstract}


\end{frontmatter}
\section{Introduction}
Fluid-acoustic shape optimization is a challenging numerical process, as both aero-hydrodynamic and acoustic responses in systems such as aircraft wings, marine propellers, and wind turbine blades are highly sensitive to shape changes \cite{portillo2024hydro, botero2023experimental}. Various high-fidelity numerical approaches such as large-eddy simulations for fluid flow combined with acoustic analogies are routinely used to obtain physical insight into the acoustic and vibration sensitivities of propeller geometry \cite{monfaredi2021unsteady,yangzhou2023aeroacoustic}, cavitation \cite{vernengo2016physics, gaggero2017efficient} and trailing edge flow kinematics with turbulent boundary layers \cite{lee2021turbulent, kou2023aeroacoustic}. These combined fluid-acoustic analyzes often require significantly different length-scale resolutions and domain sizes for fluid flow and acoustics \cite{ewert2003acoustic}. Despite advances in computational capacity, these techniques remain computationally expensive, limiting their scalability for design and optimization tasks. An alternative methodology commonly used for aeroacoustic simulations is the application of fast semi-empirical acoustic prediction models, such as Amiet’s theory \cite{amiet1976noise}, which relate noise emission to the wall pressure spectrum \cite{goody2004empirical,rozenberg2012wall,catlett2016empirical,kamruzzaman2015semi}. These semi-empirical models compute the wall pressure spectrum by integrating flow quantities across the boundary layer to predict noise levels and enable more efficient aero-acoustic shape optimization. However, high-fidelity computational fluid dynamic simulations are still required to accurately capture flow quantities across the boundary layer. This limitation motivates the development of reduced-order models for the fluid-acoustic simulations presented in this article.

In response to these limitations, much of the research has shifted towards the development of surrogate or reduced-order models that can replace traditional numerical solvers with scalable and efficient data-driven approaches \cite{moni2024data,li2019data, mallik2024deep,li2024deep,xu2020multi}. These models emulate computationally expensive full-order models and construct low-dimensional representations that overcome the computational challenges of high-fidelity models, particularly in multi-query analyses and design optimization tasks, thereby making the process more computationally efficient. Such data-driven surrogate models employ an offline-online framework: during the offline phase, the model is trained on high-dimensional physical data to capture a low-dimensional representation of the system, and in the online phase, it delivers efficient and accurate predictions. Surrogate-based shape optimization typically involves three key steps: shape parameterization, surrogate modeling, and optimization. While the primary focus of this article is on the development of surrogate models for shape optimization, it is essential to describe the shape representation and optimization process to demonstrate the overall framework.

The first important step in efficient shape optimization is geometric parameterization. The choice of parameterization method is critical to define the design space and directly influences the complexity of the optimization problem. Explicit techniques, such as the free-form deformation method \cite{sederberg1986free}, are widely used for shape optimization of airfoil and hydrofoil geometries \cite{junior2022intelligent, hui2020fast}. However, design variables (control points) in the free-form deformation method may have no physical significance to design engineers, and the large number of parameters controlling shape deformation can lead to abnormal or unrealistic shapes \cite{li2021data}, negatively impacting the efficiency of the optimization process. On the other hand, implicit shape representation via the level set methods \cite{sethian1999level,allaire2002level}, offers smoother deformations and is compatible with unstructured grids. These methods are well established for the optimization of complex topologies \cite{osher2003constructing, jiang2018parametric}, high-quality geometric modeling in computer vision \cite{park2019deepsdf} and neural network-based shape learning \cite{bhatnagar2019prediction}.

In surrogate modelling, one of the principal techniques is to project a high-dimensional dataset onto an optimal low-dimensional subspace, either linearly or nonlinearly, to reduce spatial dimension
and extract flow features. These low-dimensional analyses can provide
essential flow dynamics for design optimization.
Various techniques have been proposed for low-dimensional modeling and predictive offline-online workflows. Methods such as proper orthogonal decomposition \cite{lumley1967structure, sirovich1987turbulence} and dynamic mode decomposition  \cite{schmid2010dynamic}, along with their extended variants (e.g., \cite{towne2018spectral,schmidt2019conditional,zhang2019online}), typically project data into linear subspaces. However, these linear subspace-based methods face significant challenges in convection-dominated and high Reynolds numbers scenarios. In such cases, achieving satisfactory approximations often requires a prohibitively large number of linear subspaces.
In the literature, there are also other surrogate modelling methods such as Kriging \cite{jeong2005efficient}, polynomial response surface  \cite{queipo2005surrogate}, radial basis function \cite{forrester2009recent}, and support vector regression \cite{feng2018multidisciplinary} which have been employed for optimization applications. Among these, the Kriging or Gaussian process regression is a widely used method in various engineering analysis and design tasks \cite{han2013surrogate, hwang2018fast, mallik2020kriging}. However, Kriging suffers from a variety of issues, such as being poor at approximating discontinuous functions \cite{raissi2016deep}, difficulty in handling high-dimensional problems \cite{perdikaris2017nonlinear}, expensive to use in the presence of a large number of data samples \cite{forrester2008engineering}, and is difficult to implement for solving certain inverse problems with strong nonlinearities \cite{bonfiglio2018improving}.

In the last decade, deep neural networks have been widely adopted as alternatives to the aforementioned techniques to reduce online computational costs in shape optimization tasks \cite{li2022machine}. Machine learning algorithms, including artificial neural networks \cite{du2021rapid, bouhlel2020scalable,sekar2019fast} and convolutional neural networks \cite{miyanawala2017efficient}, have been extensively applied to shape optimization problems. Many of these ML-based surrogate models establish a direct mapping between design or operating parameters and performance metrics (such as drag or lift coefficients) through nonlinear regression. However, these models often lack a comprehensive representation of the entire flow field around the morphed geometries. Capturing the flow field enables a more detailed understanding of aerodynamic and acoustic behavior, potentially enhancing the model's applicability across diverse configurations and objectives. 
Recently, convolutional neural networks have shown significant success in predicting steady-flow fields around airfoils \cite{ du2022airfoil}, typically achieving speeds orders of magnitude faster than traditional full-order simulations.  Despite their success, convolutional neural networks are restricted to operating on uniform Cartesian grids, limiting their ability to efficiently capture varying length-scale flow features, such as large-scale far-field features and small-scale near-field features in the boundary layer \cite{lino2023current}.  This limitation becomes particularly problematic when applying convolutional neural networks for fluid-acoustic applications, where we have significantly varying length scales and domain sizes of interest. Furthermore, acoustic predictions are highly sensitive to changes in pressure fluctuations and boundary layer velocity profiles, especially when considering shape optimization. Refinement of grid resolution in these critical areas requires refining the entire flow field, which can significantly increase computational costs and make the approach inefficient. A more effective approach to overcoming these limitations is to leverage physical field information at the computational fluid dynamics grid nodes, rather than relying on pixelated representations of geometry and flow fields.

Graph neural networks (GNNs) \cite{scarselli2008graph, zhou2020graph} are a class of deep learning methods that can operate on unstructured point clouds \cite{bronstein2017geometric}. Thus, unstructured computational meshes can be easily converted into graphs, where nodes correspond to mesh points, and edges represent their connections. In recent years, GNNs have been successfully applied in various domains, where data are naturally represented with a graph structure, such as structural dynamic and static problems \cite{MATRAY2024117243,gladstone2024mesh, maurizi2022predicting}, dynamic fluid flow prediction \cite{BARWEY2023112537,lino2022multi, gao2024finite, pfaff2020learning,gao2022physics}, and  dynamic of particles \cite{ma2022fast}. These mesh-based GNN models for dynamic fluid flow prediction typically utilize an encode-process-decode architecture to predict dynamic quantities of the mesh at a new time step, based on previous time steps emulating forward Euler time discretization \cite{gao2024finite}, or solve using ODE solvers \cite{lienen2022learning}. While these models have demonstrated effectiveness in accurately and efficiently predicting dynamics of the flow field over fixed geometries, extending their application to prediction of steady-flow field predictions across a wide range of shapes and operating parameters in shape optimization problems remains an open challenge.  

In this study, we propose a novel GNN-based shape optimization framework to reduce far-field noise at the airfoil's trailing edge while maintaining aerodynamic performance. We employ an implicit shape representation approach that enables the use of completely unstructured meshes. We introduce a GNN-based surrogate model that can learn the steady-flow field around varying geometries under different operating conditions. This model also accurately resolves the boundary layer, facilitating integration with acoustic models such as Amiet's theory to predict far-field sound pressure levels. Lastly, we develop a fully integrated optimization framework that combines shape representation, GNN-based model predictions, and optimization algorithms, enabling real-time optimization of both aerodynamic performance and noise reduction. Compared to traditional methods, our framework offers significant improvements in both efficiency and accuracy, providing a more effective tool for airfoil shape optimization. 
To the best of our knowledge, no prior work has integrated GNNs with shape optimization algorithms to explore their capability for flow field simulations across a variety of configurations. The most closely related work to this research is the recent study \cite{tang2024graph}, which employed a graph-based surrogate model to optimize subsurface flow.   

The remainder of this paper is organized as follows. Section \ref{sec:data_driven_methodology} presents the data-driven methodology, including the shape representation, GNN architecture, key input features, and the optimization process. In Section \ref{sec:gnn_fluid_acoustic_implementation}, we explain the implementation of the proposed GNN framework for fluid-acoustic simulations and optimization. Section \ref{sec:results_discussion} provides detailed test case results, including flow fields, integrated flow quantities across the boundary layer, acoustic predictions, and optimization results. Finally, Section \ref{sec:concluding_remarks} summarizes our findings and provides suggestions for future work.
\section{Proposed data-driven methodology for shape optimization}\label{sec:data_driven_methodology}
The multi-objective shape optimization framework proposed in this paper follows a data-driven approach with an offline-online application strategy. During the offline phase, a GNN-based surrogate model is developed and trained to predict the flow field, with a focus on accurately resolving the boundary layer, an essential dynamic in acoustic noise generation. This surrogate model significantly reduces the computational cost of evaluating aerodynamic and acoustic performance compared to direct full-order numerical simulations.
In the online phase, the trained GNN model is integrated with the optimization algorithms to provide real-time scalable predictions for flow and noise behavior during the optimization process. This integration enables efficient exploration of airfoil shapes that can optimally balance aerodynamic efficiency and noise reduction. Our proposed framework is structured around three core steps: shape representation, GNN-based surrogate model, and the optimization procedure. Each of these steps is discussed in the following subsections.

\subsection{Shape representation}\label{subsec:shape_representation}
In this section, we employ a level-set method based on the signed distance function (SDF) to implicitly represent the geometry on an unstructured CFD grid. The SDF is widely used in CNN-based surrogate models for shape learning applications, where input is typically a snapshot of the SDF on a Cartesian grid \cite{mallik2023parametric, bhatnagar2019prediction, mallik2022deep}. However, since the GNN operates on graphs derived from unstructured CFD grids, we compute the SDF directly from these grids.

\begin{figure*}[h!]
\centering
\includegraphics[width=0.6\linewidth]{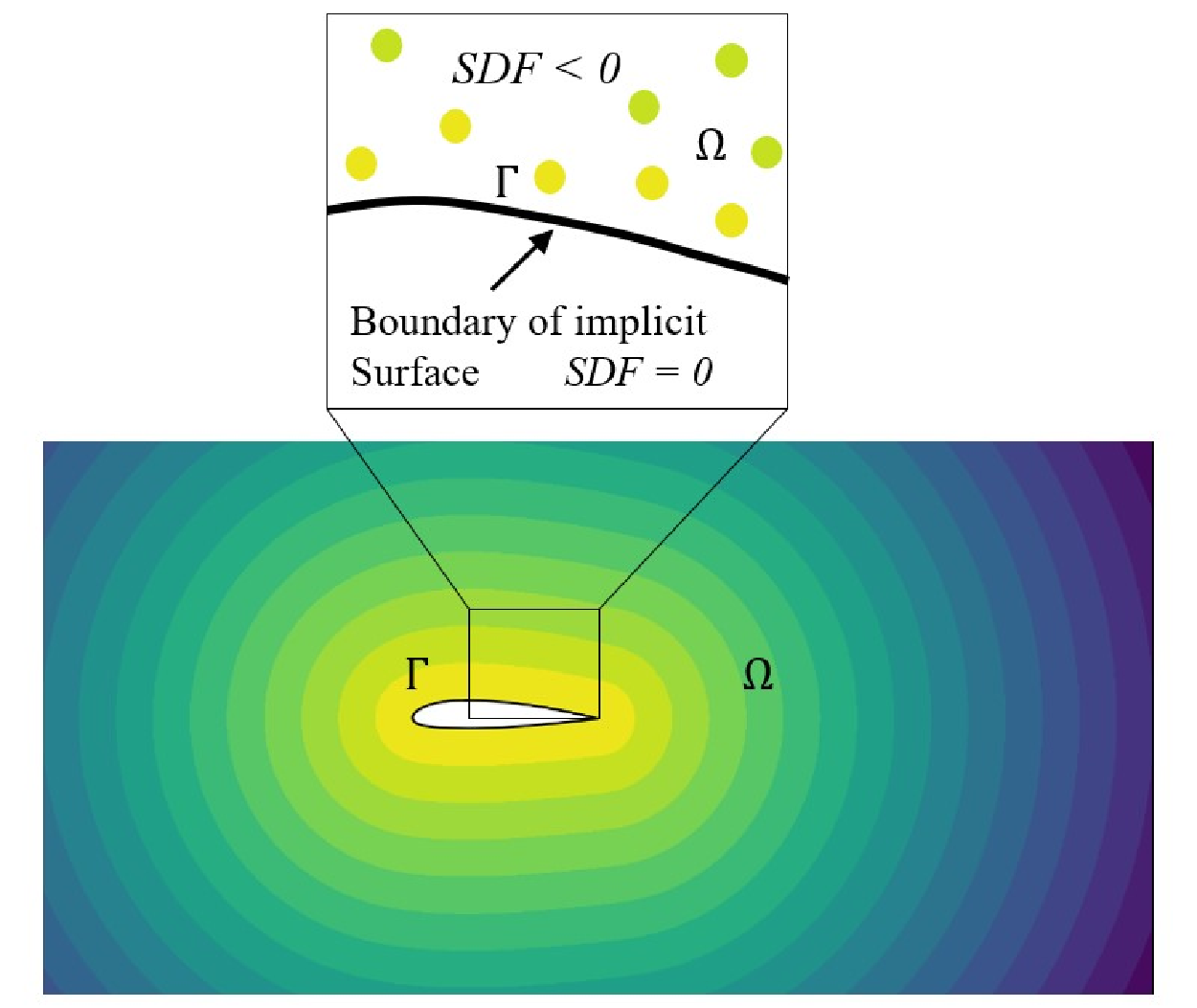}
\captionsetup{justification=centering} 
\caption{\label{sdf} Implicit representation of an arbitrary airfoil: the surface of the airfoil (\(\Gamma\)) is implicitly represented as the zero-level contour where \( SDF = 0 \). Sampled points with \( SDF < 0 \) lie outside the airfoil's boundary, within the surrounding flow field \( \Omega \).}
\end{figure*}

The SDF defines the surface of a shape as a continuous scalar field, where the magnitude at any point in the grid denotes its distance to the nearest surface boundary. The sign of the distance indicates whether the point is inside (positive) or outside (negative) the shape. The signed distance function of a set of points \( \mathbf{x} \in \Omega \subset \mathbb{R}^2\) is mathematically defined as:

\begin{equation} \label{eq:sdf}
\mathrm{SDF}(\mathbf{x}, \Gamma) =
\begin{cases}
-\min\left(\lVert \mathbf{x} - \mathbf{x}_\Gamma \rVert_2\right) & \text{for } \mathbf{x} \in \Omega, \\[0.5em]
0 & \text{for } \mathbf{x} \in \Gamma, \\[0.5em]
\min\left(\lVert \mathbf{x} - \mathbf{x}_\Gamma \rVert_2\right) & \text{for } \mathbf{x} \notin \Omega.
\end{cases}
\end{equation}
where \( \mathbf{x} \) denotes the coordinates of the grid points in the computational CFD mesh, \( \mathbf{x}_{\Gamma} \) represents the coordinates of the points on the shape boundary \( \Gamma \), and \( \Omega \) corresponds to the domain outside the boundary. Figure \ref{sdf} illustrates the SDF representation of an arbitrary airfoil, where the boundary is defined by the zero iso-surface of \( \text{SDF}(\mathbf{x}, \Gamma) = 0 \). Since the focus is on modeling the flow field outside the geometry, only the points in the domain \( \Omega \) are considered, and the region inside the shape is excluded from the computational domain. 

For each target shape, a set of pairs \( X \) is prepared, consisting of 2D point samples and their corresponding SDF values:

\begin{equation} \label{sdf-X}  
X := \left\{ \left(\mathbf{x}, s\right) : \mathrm{SDF}\left(\mathbf{x}, \Gamma\right) = s \right\}.  
\end{equation}

This set, which extracts geometric features, is used as input to the GNN model to learn geometric variations during training, as well as for flow field prediction during optimization. By representing the distance from each point on the grid to the nearest surface boundary, the SDF enables the GNN to capture spatial relationships such as proximity to geometry, sharp edges, and curvature. This feature is particularly useful for GNN to distinguish between regions near the surface, where flow behavior is highly sensitive to the geometry's concave and convex features, and regions farther away. In the following subsection, we discuss the GNN architecture and how it leverages these geometric features to model the fluid flow around the target geometry.

\subsection{GNN Framework for Shape Optimization}
\label{subsec:gnn_framework_shape_optimization}
In this section, we introduce our GNN-based surrogate model for shape optimization. The goal is to train a GNN model that predicts flow field variables, such as pressure and velocity components, from the given geometry and operating conditions during the optimization process. Thus, the proposed GNN model generates a forward map from the geometry and operating conditions to the physical flow field, as expressed in the following equation:
\begin{equation}
\begin{aligned}
\left\{\mathbf{q}^k : \left[u_x, u_y, p\right] \right\} = \mathrm{GNN}\left(\mathbf{x}\left(\Gamma\right), \mathcal{Y}; \Theta\right).
\end{aligned}  
\end{equation}
where \( \mathbf{q}^k \) represents the physical field features, including velocity components \( u_x \) and \( u_y \), as well as the pressure field \( p \). The variable \( \mathbf{x}\left( \Gamma \right) \) denotes the geometric features associated with the shape \( \Gamma \) (e.g. computational mesh and SDF representation of the geometry), and \(\mathcal{Y}\) corresponds to the operating conditions, such as angle of attack, Reynolds number, or Mach number. The parameters \( \Theta \) are the network weights learned during training.

The GNN is trained offline, independent of the optimization process. Once trained, it is integrated with an optimization algorithm to leverage its predictive capabilities efficiently. During the online inference stage, the geometry and operating parameters are provided as inputs to the GNN to predict the pressure and velocity fields. These predictions are then used to evaluate aerodynamic and acoustic performance within the optimization process. The following subsections describe the architecture of the surrogate model, the input features, and the online application procedure. Details regarding training strategies and hyperparameter explanations will be provided in Section 3.

\subsubsection{Network architecture} \label{sec:network_architecture}
\begin{figure*}[h]
\centering
\includegraphics[width=1\linewidth]{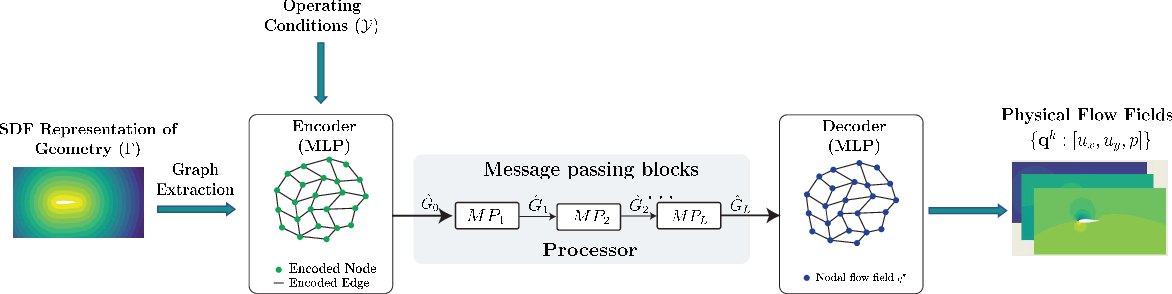}
\caption{Schematic of the model architecture, consisting of an encoder, processor, and decoder. The encoder and decoder are implemented as MLPs, and the processor is a message-passing neural network with 
\( L \) message-passing graph layers.}
 \label{overall-GNN}
\end{figure*}

The overall architecture of the proposed network is shown in Figure \ref{overall-GNN}. This architecture follows the encode-process-decode framework \cite{battaglia2018relational}, which consists of an encoder, a sequence of message-passing layers, and a decoder. As a reminder, the goal is to predict a physical field image 
\( \mathbf{q}^k : \left[ u_x, u_y, p \right]
 \) associated with the geometry \( \Gamma \), and operating parameters \( \mathcal{Y}\). Specifically, the model predicts the output \( \mathbf{q} \in \mathbb{R}^k
 \) for each node \( i \) in the computational mesh. 

The input to our proposed GNN contains information on the geometric and operating characteristics, such as the position of the nodes and the type of boundary conditions to which they are subjected. During training, the predicted output \( \hat{\mathbf{q}}^k
 \) is compared with the ground truth \( \mathbf{q}^k
 \), and the GNN weights are adjusted accordingly. During the optimization phase (online phase), the physical field \( \mathbf{q}^k_{\text{new}}
 \) is predicted for a new geometry \( \Gamma_{\text{new}}
 \) or a new set of operating conditions \( \mathcal{Y}_{\text{new}}
 \). Algorithm 1 summarizes the GNN methodology for shape optimization applications. Details on the node and edge features, as well as the training strategy, are provided in the following subsections.

\RestyleAlgo{ruled}
\SetKwComment{Comment}{/* }{ */}

\SetKwInput{KwInputs}{Inputs}  
\SetKwInput{KwOutput}{Output} 
\SetKwInput{KwNetwork}{Network} 

\SetKwFor{Flower}{Offline phase (Training)}{}{\text{return \small{Trained \textit{GNN}} model}}
\SetKwFor{Sea}{Online phase (Inference)}{}{\textbf{return $\hat{\mathbf{q}}^k_{New}$}} 

\SetAlgoNlRelativeSize{0}  
\SetAlgoVlined
\SetAlgoNlRelativeSize{1}

\begin{algorithm}[hbt!]
\caption{Proposed graph neural network procedure for shape optimization}\label{alg:two}

\Flower{}{
    \KwInputs{\parbox[t]{13.8cm}{
        Dataset \(\mathcal{D} : \left\{ \Gamma_n, \mathcal{Y}_n, \mathbf{q}^k_n \right\}_{n=1, \dots, N}^{\text{GT}}
\) (Known geometries, and corresponding operating conditions, and physical fields),
        Batch size: \(\mathcal{N}_b\), Epochs: \(\mathcal{N}_{ep}\), Patience: \(\mathcal{N}_p\), Message passing layers: \( \mathcal{N}_l \),}}

    \KwOutput{Trained GNN model: \(\theta^*\)}

    \KwNetwork{}{}
    Initialize GNN parameters: \(\theta \),\\
    Initialize the counter for early stopping: \(\left( l \gets 0 \right)
\)\\
    
    \While{\( \text{epoch} < \mathcal{N}_{\text{ep}}  \text{ and } l < \mathcal{N}_{\text{p}} 
\)}{ 
        Randomly sample batch of input dataset \(\mathcal{D}\);\\
        \textbf{Encoding:} encode nodes and edges: \(\hat{G}_0\left( E, V \right)\)\\
        \(h = 0;\)\\
        
        \While{\(h < \mathcal{N}_l \)
}{
            \(e^{\prime}_{ij} \leftarrow \phi_e\left( v_i, v_j, e_{ij} \right)\)
\\
            \(v^{\prime}_i \leftarrow \phi_v\left( v_i, \text{AGG}\left( e^{\prime}_{ij} \right) \right)
\)\\
            \(h \leftarrow h + 1
\)
        }
        
        \textbf{Decoding:} \(\left\{\hat{\mathbf{q}}^k_i : \left[u_x, u_y, p\right]_i \right\}_{i=1, \dots, n_n} = h_v\left( v^{\prime}_i \right)
\)\\
        
        Calculate loss \(\mathcal{L}\) from Eq. \ref{eq:loss} \\ 
        Update parameters \(\theta \leftarrow \text{ADAM}\left( \hat{g} \right)
\)\\
        
        \eIf{\(\mathcal{L} > l_b\)}{
            Save the model parameter \;
        }{
            \(l \leftarrow l + 1
\)
        }
    }
}
\vspace{0.5cm}

\Sea{}{
    \KwInputs{New geometry \( \Gamma_{\text{new}}
\), and new operating conditions \(\mathcal{Y}_{\text{new}}\) (unknown to the GNN), optimized GNN parameters: \(\theta^{*}
\)}

    \KwOutput{Physical flow fields for the new geometry and operating condition: \(\hat{\mathbf{q}}^k_{\text{new}}
\)}

    \KwNetwork{}{}
\(   \hat{\mathbf{q}}^k_{\text{New}} : [u_x, u_y, p]_{\text{new}} = \text{GNN}(\Gamma_{\text{new}}, \mathcal{Y}_{\text{new}})
\)\\
}

\end{algorithm}

Once the graph is generated from the computational mesh \( \mathcal{M} \) for the geometry \( \Gamma \), the node and edge features are encoded into a latent space through an encoder. The encoder performs encoding for each node and edge of the graph and is composed of two parts: the node encoder, and the edge encoder. These encoders transform the computational mesh state \( L \) into the graph input \( \hat{\mathrm{G}}_0\left( E, V \right)
 \), where \( \hat{\mathrm{G}}_0
 \) is the encoded graph input, \( V \) is the set of all nodes and \( E \) is the set of all edges. Each node is associated with a reference mesh-space coordinate and the physical quantities to be modeled (e.g., pressure and velocity components). Both encoders utilize multilayer perceptrons for this transformation.

Next, the processor, which is a message-passing neural network consisting of \( L \) identical message-passing layers, facilitates information exchange across the graph. The value of \( L \) determines the set of neighboring nodes from which the GNN gathers information. For each message-passing layer \( i \) (where \( i = 1, \dots, L
 \)), the input is the output of the previous layer, denoted as \( \hat{\mathrm{G}}_{i-1}
 \), and the output is the updated graph \( \hat{\mathrm{G}}_i
 \). The final encoded graph after \( L \)  steps of the message passing layers is denoted \( \hat{\mathrm{G}}_L
 \). Each message passing layer updates node and edge features through a two-stage process. In the first stage, the edge features \( e_{ij} \) are updated as follows:

\begin{equation}
\begin{aligned}
e^{\prime}_{ij} &\leftarrow \mathrm{\phi_e}\left( v_i, v_j, e_{ij} \right),
\end{aligned}  
\end{equation}
where \( \phi_e \) denotes the edge update function, which is a multilayer perceptron. \( e^{\prime}_{ij} \) denotes the updated edge features after one message passing step. \( v_i \) and \( v_j \) denote the features of the sender node and receiver node of the edge, respectively.

In the second stage, the updated edge features are used as messages and are aggregated using an aggregation function \( \text{AGG} \) at the receiver of each edge. These aggregated features are then concatenated with the receiver's node input \( v_i \). The concatenated features are passed into a node update function \( \phi_v \), which is also a multilayer perceptron with the latent dimension of \( h \), as shown below:
\begin{equation}
\begin{aligned}
v^{\prime}_i &\leftarrow \mathrm{\phi_v}\left( v_i, \mathrm{AGG}\left( e^{\prime}_{ij} \right) \right),
\end{aligned}  
\end{equation}
The above stages are combined into a single message-passing layer: 
\begin{equation}
\begin{aligned} \label{message_update}
v^{\prime}_i, e^{\prime}_{ij} &= \hat{G}_{n}\left( v_{i}, v_{j}, e_{ij} \right), \hspace{1.5cm} n = 1, \dots, L.
\end{aligned}  
\end{equation}
Finally, the decoder, denoted as \( h_v \), reverses the encoding process. This decoder uses a multilayer perceptron with two hidden layers 
to transform the latent node features \( \hat{G}_{L} \) of dimension \( h \), derived after the final processing step, into the output features of dimension \( k \).
\begin{equation}
\begin{aligned}
\left\{ \hat{\mathbf{q}}^k_i : \left[ u_x, u_y, p \right]_i \right\}_{i=1, \dots, n} &= h_v\left( v^{\prime}_i \right).
\end{aligned}  
\end{equation}
in which \( \hat{\mathbf{q}}^k_i\) contains the state variables \( k \) (such as pressure or velocity components) corresponding to the given geometry and operating conditions. Here, \( n \) is the number of nodes in the computational grid.

\subsubsection{Graph neural network features} \label{sec:gnn_features}
The GNN operates on graph data consisting of nodes and edges, so the features are categorized into node and edge features. The input to the node encoder includes \( \mathrm{SDF}_i \), the signed distance representation of the \( i \)-th node, flow operating features \( \mathcal{Y} \), and a one-hot node type vector \( \gamma_i \), which defines the node type and differentiates between nodes in the fluid domain and various boundary types, such as wall surfaces, inlets, and outlets.
\begin{equation}
\begin{aligned} \label{node_features}
v_i = \left[ \mathrm{SDF}_i, \gamma_i, \mathcal{Y} \right].
\end{aligned}  
\end{equation}

The input to the edge encoder consists of the Euclidean distance between the nodes forming each edge, along with the edge length, denoted as:
\begin{equation}
\begin{aligned} \label{edge_features}
e_{ij} = \left[ x_i - x_j, y_i - y_j, l_i \right]^T.
\end{aligned}  
\end{equation}
where \( l_i = \sqrt{\left( x_i - x_j \right)^2 + \left( y_i - y_j \right)^2}
 \) is the length of the edge, and \( j \) denotes the index of the node connected to the \( i \)-th edge.
At the end of the offline phase described in this section, a GNN model is obtained that predicts the physical field \( \mathbf{q}^k_{\text{new}}
 \) for a new, arbitrary geometry \( \Gamma_{\text{new}} \). As detailed in the following section, this \( \mathbf{q}^k_{\text{new}}
 \) can then be used to obtain the aerodynamic and acoustic performance metrics needed for the optimization phase.

\subsection{Optimization process} \label{sec:optimization_process}

For online applications during shape optimization, the trained GNN surrogate model is integrated into an evolutionary optimization framework. This integration enables the rapid evaluation of objective functions, which is important for efficiently exploring the design space. The objective of the shape optimization process is to determine the location of the shape boundaries \( \Gamma \) for a given set of operating conditions \( \mathcal{Y} \), such that one or more objective functions are maximized or minimized. The multi-objective optimization problem is stated as:

\begin{align}
    S^* &= \text{argmin} \ \left\{\mathcal{J}_{1}(S, \mathcal{Y}), \mathcal{J}_{2}(S, \mathcal{Y}), \dots, \mathcal{J}_{k}(S, \mathcal{Y})\right\} \notag \\
    &\begin{aligned}
    \qquad \text{subject to:} \quad & g_{j}(S, \mathcal{Y}) \leq 0, \hspace{1cm} j = 1, 2, \dots, J, \\
                              & S_{i}^{L} \leq S_{i} \leq S_{i}^{U}, \hspace{0.8cm} i = 1, 2, \dots, I.
    \end{aligned}
\end{align}
where \( \mathcal{J}_k(S, \mathcal{Y}) \) represent the k-th objective functions to be optimized, \( S \) denotes the geometric design variables which defines the airfoil profile \( \Gamma \), and \( \mathcal{Y} \) represents the operating conditions. \( S^* \) represents the optimal geometric design variables that define the optimal geometry profile \( \Gamma^* \), and \( g_j(S, \mathcal{Y}) \) are nonlinear constraints that may include geometric or simulation-related quantities.

In each iteration, the optimization process begins with the initialization of geometric design variables, which are iteratively updated by the optimization algorithm. For each candidate solution, the geometric parameters are proposed by the optimization algorithm. The geometry is then represented using the signed distance function and the node and edge features are constructed as described above. The trained GNN model uses these features to predict the system state \( \mathbf{q}^k \), including physical quantities such as pressure and velocity, for candidate designs generated by the optimization algorithm. These predictions are then used to evaluate the objective functions \( \mathcal{J}_k(S, \mathcal{Y})\). At each generation, the optimization algorithm ensures that constraints \( g_j(S, \mathcal{Y}) \) are satisfied. If necessary, the design variables are refined for the next generation. Based on these evaluations, the optimization algorithm selects and updates the design variables for subsequent iterations. This iterative process continues until convergence is achieved or a predefined number of iterations or generations is reached. Further details about the integration of the GNN with the optimization framework are provided in Section~\ref{sec:GNN_integration}.  

\section{Implementation of GNN for fluid-acoustic shape optimization} \label{sec:gnn_fluid_acoustic_implementation}
To assess the performance of our proposed methodology, we consider a case study on acoustic-fluid shape optimization for 2D airfoils. Previous studies \cite{lutz2007design} have shown that trailing-edge noise is the dominant source of noise for airfoils. This noise comes primarily from the interaction of turbulent eddies within the boundary layer and the associated pressure fluctuations near the trailing edge of the airfoil. Since the state of the turbulence in this region is determined by the boundary layer development and the pressure distribution along the blade section, it is clear that noise emission can be influenced—and ultimately reduced—by optimizing the airfoil shape. Specifically, the thickness of the airfoil and the leading-edge radius directly affect the turbulent inflow noise \cite{hao2008aerodynamic}. In this context, we propose a multi-objective shape optimization framework to identify the optimal airfoil geometry that maximizes aerodynamic performance while minimizing trailing-edge noise generation. The key idea is to employ our GNN as a surrogate model to alleviate the computational costs associated with the shape optimization process.

As outlined earlier, our optimization framework follows an online-offline strategy. In the offline phase, we consider a set of \( N \) independent geometries and operating conditions \( \left\{ \Gamma_{n}, \mathcal{Y}_n \right\}^{\text{GT}}_{n=\{1, \dots, N\}}
\), each associated with a full-order solution \( \left\{ \mathbf{q}^k_n \right\}^{\text{GT}}_{n=\{1, \dots, N\}}
 \) obtained through CFD simulations. We then train the GNN using supervised learning in this dataset to predict \( \mathbf{q}^k_{\text{new}}
 \) for new geometries \( \Gamma_{\text{new}}
\) and operating conditions \(\mathcal{Y}_{\text{new}}
 \)during the optimization phase. Aerodynamic and acoustic performances are derived from the predicted \( \hat{\mathbf{q}}^k_{\text{new}}
 \), analogous to the CFD solution.

In the following subsections, we first apply our GNN-based surrogate model (developed in Section \ref{subsec:gnn_framework_shape_optimization}) to predict the behavior of fluid and acoustic systems in 2D airfoils. Upon completion of the training, the model will be integrated with an optimization algorithm to identify the optimal shapes.

\subsection{GNN training for fluid simulation} \label{subsec:gnn_training_fluid_simulation}
In this section, we apply the GNN-based surrogate model developed in Section 2.2 to predict the fluid flow around 2D airfoils as a proof of concept. We begin by describing the dataset and the necessary preprocessing steps before transforming the CFD mesh into the GNN features. We then provide details on the network setup, implementation, and training processes. Finally, we present the approach used to quantify the error.

\subsubsection{Dataset} \label{subsubsec:dataset}
In this work, we use the AIRFRANS dataset \cite{bonnet2022airfrans} to train and evaluate the performance of our GNN model. This dataset provides  \( N=1000 \) cases of incompressible steady-state Reynolds Average Navier-Stokes (RANS) simulations in two dimensions, including a wide range of NACA 4 and 5-digit series, various Reynolds numbers and angles of attack. Each simulation is defined by an airfoil geometry profile, a Reynolds number (ranging from 2 to 6 million), and an angle of attack (ranging from -5 to 15 degrees). 

This dataset is divided into training, validation, and test sets. Specifically, 70\% of the cases are allocated to the training set, where the GNN adjusts its parameter weights during the learning process. This subset consists of \( \left\{ \Gamma_{n}, \mathcal{Y}_n, \mathbf{q}^k_n \right\}^{\text{GT}}_{n=\{1, \dots, N=700\}} \), representing various geometries, Reynolds numbers, and angles of attack. Another 10\% is designated as the validation set, which includes cases that the network has not encountered during training. These cases are used to fine-tune the model and mitigate overfitting by employing early-stopping criteria \cite{rice2020overfitting}. The remaining 20\% forms the test set, which contains entirely new combinations of geometries and operating conditions \( \left\{\Gamma_\text{new}, \mathcal{Y}_\text{new} \right\} \). These cases, which are unknown to the GNN model, are used to evaluate its performance after training.

Full-order simulations \(\left\{ \mathbf{q}^k_n \right\}^{\text{GT}}_{n=\{1, \dots, N\}}
\) were performed using the OpenFOAM solver \cite{weller1998tensorial} with the \(k-\omega\) SST turbulence model, under sea-level conditions and a temperature of 298.15 K. The flow fields for each case, including the pressure \( p \) and velocity components \( u_x \) and \( u_y \), were computed using the steady-state RANS solver simpleFOAM. A computational mesh is generated for each airfoil profile using the blockMesh utility, with mesh refinement applied near the airfoil surface to ensure that \( Y^+ \) remains below 1. Further technical details, including the meshing procedure, boundary conditions, and RANS equations, can be found in \cite{bonnet2022airfrans}. Table \ref{tab:dataset} summarizes the range of parameters used in this dataset and in our training.

\begin{table}[h]
    \centering
    \captionsetup{font=small,  justification=centering}
    \caption{Summary of parameters provided by the AIRFRANS dataset, used for training, validation, and testing of our GNN-based surrogate model.}
    \label{tab:dataset}
    \small
    \setlength{\tabcolsep}{4pt}
\renewcommand{\arraystretch}{1.3}
    \begin{tabular}{c c c c c c c c c c c c}
        \toprule[1pt]
        \multirow{2}{*}{\makecell{Dataset}} & \multirow{2}{*}{\makecell{Number of \\ Cases}} & \multirow{2}{*}{\makecell{Angle of \\ Attack}} & \multirow{2}{*}{\makecell{Reynolds \\ Number}} & \multicolumn{8}{c}{NACA 4 and 5-digit parameters}  \\
        \cmidrule{5-7} \cmidrule{8-12}
        & & & & \( M \) & \( P \) & \( XX \) & & \( L \) & \( P \) & \( S \) & \( TT \) \\
        \midrule
        Airfoil & 1000 & \(-5^\circ - 15^\circ\) & \(2 \times 10^6 - 6 \times 10^6 \)& \( [0,7] \)& \(\left\{0\right\} \cup [1.5,7] \)& [5,20] &  & [0,4] & [3,8] & \(\left\{ 0,1 \right\}\) & [5,20] \\
        \bottomrule[1pt]
    \end{tabular}
\end{table}

\subsubsection{Interpolation on neural network grid} \label{subsubsec:interp_nn_grid}
Training graph neural networks using node and edge features derived from large CFD computational grids is computationally expensive. To reduce the numerical complexity of the problem, we interpolate full-order spatial fields \( \left\{\mathbf{q}^k : \left[u_x, u_y, p\right] \right\} \) from the fine computational grid to a coarser neural network grid suitable for GNN training. This interpolation is performed using a Linear Interpolator available in the SciPy package \cite{virtanen2020scipy}, and mathematically expressed as:
\begin{equation}
\mathbf{q}_{\text{coarse}}^k = \mathcal{I}(\mathbf{q}_{\text{fine}}^k),
\end{equation}
where \( \mathcal{I} \) represents the interpolation operator. Given a full-order CFD solution field \(\mathbf{q}^k(\mathbf{x})\) defined at fine-grid points \(\mathbf{x}_i\) (\(i = 1, \dots, N_{\text{fine}}\)), the interpolated field \(\mathbf{q}_{coarse}^k(\mathbf{z}_j)\) at coarse-grid points \(\mathbf{z}_j\) (\(j = 1, \dots, N_{\text{coarse}}\)) is computed as:
\begin{equation}
\mathbf{q}_{coarse}^k(\mathbf{z}_j) = \sum_{i \in \mathcal{A}(\mathbf{z}_j)} w_{ij} \mathbf{q}^k (\mathbf{x}_i).
\end{equation}
where \( \mathcal{A}(\mathbf{z}_j) \) represents the set of neighboring points in the fine grid contributing to the interpolation, and \( w_{ij} \) are the interpolation weights determined based on the interpolation method.

\begin{figure*}[h]
    \centering
    \includegraphics[width=0.6\linewidth]{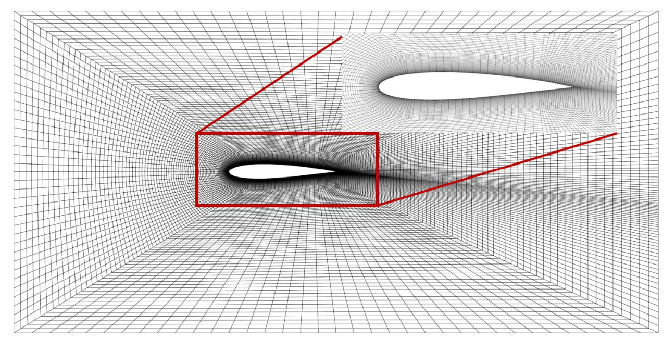}
    \captionsetup{font=small}
    \caption{Schematic of a sample mesh used for training our GNN-surrogate model, with high resolution in the shearing layer.}
    \label{fig:mesh}
\end{figure*}

Since grid resolution directly impacts prediction accuracy, a grid independence study is conducted to determine the optimal mesh resolution. This study ensures that further grid refinements do not significantly affect aerodynamic and acoustic performance metrics. The coarse computational mesh used for the GNN training is generated using Gmsh \cite{geuzaine2009gmsh}, with a higher resolution in critical regions (e.g., wall surfaces and wake zones) and sparser resolution in less sensitive areas (Figure~\ref{fig:mesh}). The aerodynamic and acoustic performance indicators—specifically the lift coefficient \( C_L \) and the overall A-weighted sound pressure level (OASPL)—are used as evaluation metrics. Details on calculating overall A-weighted sound pressure level are provided in subsection \ref{subsec:Far-field noise prediction}.  

The sensitivity of \( C_L \) and overall A-weighted sound pressure level to grid resolution, as shown in Table \ref{tab:grid}, becomes minimal when the grid numbers exceed 16,616. Thus, this resolution is used for all subsequent calculations. After determining the optimal grid resolution, the coarse computational meshes for all cases in the dataset are generated using an automated mesh generation strategy controlled through a Python API. This automated strategy ensures that Gmsh generates consistent, high-quality meshes for each geometry \( \left\{ \Gamma_{n}\right\}^{\text{GT}}_{n=\{1, \dots, N=1000\}} \)across all cases in the dataset such that the number of cells, nodes, and connectivity remain identical for all cases. After interpolation across all the cases in the dataset onto the coarser neural network grid, the resulting spatial fields \( \left\{ \mathbf{q}^k_n \right\}^{\text{GT}}_{n=\{1, \dots, N=1000\}} \) are transformed into the geometric node and edge features for GNN training.

\begin{table}[h]
    \centering
    \captionsetup{font=small, skip=6pt, justification=centering}
    \caption{Grid Independence Study conducted to identify the coarse grid used for training our proposed GNN-based surrogate model: Effect of Mesh Resolution on the Overall A-weighted Sound Pressure Level and Lift Coefficient. The grid numbers represent the total number of cells in the mesh.}
    \label{tab:grid}
    \footnotesize
    \small
    \setlength{\tabcolsep}{10pt} \renewcommand{\arraystretch}{1.05}
    \begin{tabular}{c c c c c}
        \toprule[1pt]
        & \multicolumn{4}{c}{Grid Number} \\
        \cmidrule{2-5}
        & 3552 & 7674 & 16616 & 24440 \\
        \midrule
        OASPL (dB) & 156.57 & 158.12 & 157.94 & 156.72 \\
        $C_L$ & 0.2446 & 0.2561 & 0.2666 & 0.2683 \\
        \bottomrule[1pt]
    \end{tabular}
\end{table}
\subsubsection{Network setup, implementation, and training} \label{subsubsec:network_setup}
We now describe the procedures used to train our GNN to predict static fluid flow states. Our model is implemented using the PyTorch framework \cite{paszke2019pytorch}. As mentioned in previous sections, all multi-layer perceptrons  in the network—encoders, decoders, and processors—have two hidden layers with a width of 128, except for the output layer of the decoders, which matches the size of the output predictions. All MLP outputs, except for those of the decoder’s MLP, are normalized using LayerNorm, and the ReLU activation function is employed for all hidden and output layers. We utilize a total of 15 message-passing layers. It is important to note that the hyperparameters of network width and depth are not tuned but are fixed, following the choices of Pfaff et al. \cite{pfaff2020learning}. The sum aggregation function is employed during both the edge and node update stages.

Additionally, the input matrix \( \mathbf{I} \in \mathbb{R}^{n \times r} \) to the GNN includes nodal SDF values, node type (boundary condition type), and operating conditions (Reynolds number and angle of attack). The output matrix of the GNN, \( \mathbf{q}^k \in \mathbb{R}^{n \times k} \), consists of flow field states, specifically pressure and velocity components. Thus, at each node, there are \( r = 7 \) elements of input and \( k = 3 \) elements of output. GNN input data are pre-processed using standard normalization techniques to improve learning performance. Input, output and network features are normalized using zero-mean and unit-standard deviation scaling based on the entire training dataset, with a minimum of 0 and a maximum of 1. The test datasets are normalized using the statistics calculated from the training datasets. To enhance model robustness against noisy input, we add Gaussian noise to the ground truth input and output values during training. Specifically, Gaussian noise with a mean of 0 and a standard deviation of 0.001 is added to the normalized training data, leading to the perturbed quantities \( \tilde{\mathbf{q}}^k = \mathbf{q}^k_{norm} + \mathrm{\Psi}
 \), where \( \mathbf{q}^k_{norm}
 \) is the normalized data and \( \Psi \sim \mathcal{N}(0, \sigma^2)
 \) represents the Gaussian noise, with \( \sigma \) denoting the standard deviations.

We utilize both mean squared error (MSE) and mean absolute error (MAE) in the loss expression. While MSE is commonly applied in deep learning models, the contribution of MAE is beneficial as it reduces the impact of outliers and provides a constant gradient, which can facilitate training. The resulting loss function \( \mathcal{L} \) that we seek to minimize is expressed as

\begin{equation}
\label{eq:loss}
\mathcal{L} = \frac{1}{n_s n_n} \left( \sum_{i=1}^{n_s} \sum_{j=1}^{n} \| \hat{\tilde{q}}^k_{i,j} - \tilde{q}^k_{i,j} \|_2^2 + \beta \sum_{i=1}^{n_s} \sum_{j=1}^{n} \| \hat{\tilde{q}}^k_{i,j} - \tilde{q}^k_{i,j} \| \right).
\end{equation}
where \( n_s \) is the number of samples used in training, \( n \) is the number of nodes, and \( \beta \) is the ratio of the MAE loss to MSE loss for matrix nodes, which is chosen as 0.1 after experiments. We use the notation \( \hat{\mathbf{\tilde{q}}}^k_{i,j}
 \) to indicate the prediction for state variable \( k \) \(\left( u_x, u_y, p \right)
\) for node \( j \) in sample \( i \), and \( \mathbf{\tilde{q}}^k_{i,j}
 \) for the corresponding perturbed ground truth data. 
 
The model is trained using Adam Optimizer via the stochastic gradient descent algorithm \cite{kingma2014adam} with PyTorch's default setup for a batch size of 4, and \( L_2 \) regularization applied to GNN weights via the "weight decay" parameter \( \gamma \). We implement exponential learning rate decay, starting from \(10^{-4}\) and gradually decreasing to \(10^{-6}\) over the course of 3000 epochs. Additionally, we use early stopping to stop the training if there is no progress on the validation set. A patience of 200 epochs is set, meaning that if the validation mean squared error does not decrease within this period, the training process will stop. The hyperparameters associated with the best model are provided in Table \ref{tab:hyper}. The training time required for the network is approximately 40 hours using a single Nvidia A100 GPU without parallelization. 

\begin{table}[h]
    \centering
    \captionsetup{font=small, justification=centering}
    \caption{Summary of the hyperparameters used in the training of the proposed GNN surrogate model.}
    \label{tab:hyper}
    \small
    \setlength{\tabcolsep}{4pt}
    \renewcommand{\arraystretch}{1.3}
    \begin{tabular}{l l l} 
        \toprule[1pt]
        No. & Hyperparameter & Value  \\
        \midrule
        1. & Number of hidden layers & 2 \\
        2. & Hidden size & 128 \\
        3. & Number of message passing layers & 15 \\
        4. & Type of message passing & neighbor info \\
        5. & Type of aggregation function & summation \\
        6. & Types of activation & ReLU \\
        7. & Gaussian noise std. dev. & 0.001 \\
        8. & MAE loss ratio & 0.1 \\
        9. & Learning rate & \(10^{-4}-10^{-6}\) \\
        \bottomrule[1pt]
    \end{tabular}
\end{table}

\subsubsection{Evaluation metrics} \label{subsubsec:evaluation_metrics}
The main purpose of training surrogate models for optimization applications is to accurately predict the system states  \(\mathbf{q}^k\)  over unseen shapes and operational conditions during the optimization process. We evaluate the model's performance by feeding the geometric and operational parameters from each of the test datasets to the model and comparing the predicted system states with the ground truth values from the interpolated CFD data.

We will present test-case results in terms of relative error statistics. The relative error for system states (pressure and velocity components) for test sample \( i \), denoted as \( e_p^i \) and \( e_u^i \), is given by:

\begin{equation}
\begin{aligned} \label{error:estate}
	e^i_p &= \frac{1}{n} \sum_{j=1}^{n} \frac{\left| \hat{p}_{i,j} - p_{i,j} \right|}{\frac{1}{2} \rho U_\infty^2}, \hspace{2cm} 
	&e^i_u &= \frac{1}{n} \sum_{j=1}^{n} \frac{\left| \hat{u}_{i,j} - u_{i,j} \right|}{U_\infty},
\end{aligned}
\end{equation}
where \( n \) is the number of nodes in the model, \( \hat{p}_{i,j} \) and \( p_{i,j} \) represent the pressures predicted by the GNN and obtained from the CFD simulation, respectively, for node \( j \) in test sample \( i \). Given the wide range of pressure distribution across the test cases, we normalize the pressure error using the dynamic pressure \( q = \frac{1}{2} \rho U_\infty^2
 \), where \( \rho \) is the fluid density and \( U_\infty \) is the free-stream velocity of sample \( i \). The quantities \( \hat{u}_{i,j} \) and \( u_{i,j} \) denote the velocity values from the GNN and the CFD simulation, respectively, for node \( j \) in test sample \( i \). For the velocity error, since the velocity \( u \) at the surface of the geometry is zero, we normalize by the free-stream velocity \( U_\infty \).
Relative error for integrated quantities \(\mathcal{Q}\), \(e^i_{\mathcal{Q}}\), is given by
\begin{equation}
\begin{aligned} \label{error:integ}
    e^i_{\mathcal{Q}} = \frac{\left|\hat{\mathcal{Q}}_{i}-{\mathcal{Q}_{i}}\right|}{\mathcal{Q}_{i}}.
\end{aligned}
\end{equation}
where \( \hat{\mathcal{Q}}_{i} \) and \( {\mathcal{Q}_{i}} \) are the integrated quantities obtained from the GNN prediction and the CFD simulation for test sample \( i \), respectively. The integrated quantities \( \mathcal{Q} \) could be either the aerodynamic coefficients \( C_L \) or \( C_D \), or the integrated boundary layer parameters, such as displacement thickness \( \delta^* \) and momentum thickness \( \theta \).

 \subsection{Far-field noise prediction }\label{subsec:Far-field noise prediction}
The flow-field predictions from the GNN model are integrated with an acoustic model to predict trailing-edge noise generated by turbulent boundary layers on 2D airfoils. The acoustic model used in this work is based on Amiet's theory \cite{amiet1976noise}. Amiet's theory calculates the far-field noise by relating the wall pressure wavenumber–frequency spectral density, \(\Pi_{pp}\left(k_{x}, k_{z}, \omega\right)
\), and the radiated far-field power spectral density of the acoustic pressure, \( S_{pp}\left(x, y, z = 0, \omega\right) \), at the trailing edge. Under the assumptions of a large span, a stationary observer and airfoil, and a uniform free-stream flow, the general expression for \(S_{pp}
\) at frequency \( \omega \) is given by \cite{stalnov2016towards}:

\begin{equation}
\begin{aligned} \label{spp}
S_{pp}(x_{0}, y_{0}, z_{0}, \omega) = \left( \frac{\omega c y_{0}}{4 \pi c_{0} \sigma^2} \right)^2 \frac{b}{2} \left| \mathscr{L}\left( \frac{\omega}{\overline{U}_{c}}, k_{z} = 0, x, y, U_{\infty}, \overline{U}_{c} \right) \right|^2 \Lambda_{z | pp} \Pi_{pp}\left( \frac{\omega}{\overline{U}_{c}}, k_{z} = 0, \omega \right),
\end{aligned}  
\end{equation}
where \( c_0 \), \( c \), and \( b \) are the speed of sound, the chord length, and the span of the airfoil, respectively. \( \omega = 2 \pi f \) is the angular frequency, and \( k_x \) and \( k_y \) represent the streamwise and spanwise wavenumbers.

The observer's location is given by \( x_0 \), \( y_0 \), \( z_0 \), with the origin of the coordinate system at the trailing edge. In this system, \( x \) is in the streamwise direction, \( y \) is in the wall-normal direction, and \( z \) is in the spanwise direction. \( U_\infty \) is the freestream velocity, and \( \overline{U}_c \) is the mean effective convection velocity, which represents the net averaged speed of eddies of different sizes in the boundary layer. \( \sigma \) is the flow-corrected radial distance, \( \sigma^2 = x_{0}^2 + \beta^2 \left( y_{0}^2 + z_{0}^2 \right)
 \), where \( \beta^2 = 1 - M^2
 \), and \( M \) is the Mach number.

The term \( \mathscr{L} \) is the aeroacoustic transfer function that represents the relationship between wall pressure fluctuations caused by the turbulent boundary layer and the acoustic waves in the far field. We have used the \( \mathscr{L} \) proposed by Roger and Moreau \cite{roger2005back}, which considers both the backscattering effect of the leading edge and the scattering caused by the trailing edge.  \( \Lambda_{z|pp}
 \) is the spanwise correlation length scale of the inflow turbulence, and \( \Pi_{pp}
 \) is the wall pressure frequency spectrum close to the trailing edge. 

To compute the terms in the power spectral density equation Eq. \ref{spp}, we extract the boundary layer parameters from the flow field predicted by the trained GNN model, for the target geometry and operating conditions. These parameters include boundary layer thickness \( \delta \), displacement thickness \( \delta^* \), momentum thickness \( \theta \), shear stress distribution, and pressure gradients—all of which significantly influence the acoustic sources near the trailing edge. Detailed procedures for calculating the terms in Eq. \ref{spp} are provided in \ref{sec:Appendix-A}. 

Using these parameters, we first calculate the radiated far-field power spectral density \( S_{pp} \) from Eq. \ref{spp}. Subsequently, the sound pressure level (SPL) is determined using Eq. \ref{spp_log}: 

\begin{equation}
\label{spp_log}
SPL = 20 \log_{10} \left( \frac{S_{pp}(f)}{p_{\text{ref}}^2} \right).
\end{equation}
where \( p_{\text{ref}} \) is the reference acoustic pressure, which is \( 20 \mu Pa \) for air. finally, we apply the A-weighting filter across the entire frequency spectrum \cite{doolan2012wind} to obtain the overall A-weighted sound pressure level. The A-weighting filter adjusts the frequency spectrum by emphasizing frequencies within the 3–6 kHz range, where the human ear is most sensitive, and attenuating very high and low frequencies that are less significant. Finally, we integrate the results over the frequency range from 100 to 10 kHz.

\subsection{Integration of GNN with optimization process}
\label{sec:GNN_integration}
Now we integrate our trained GNN model and acoustic model with an optimization algorithm to minimize trailing-edge noise and maximize the aerodynamic performance of 2D airfoils. The overall A-weighted sound pressure level is chosen as the acoustic objective function, while the lift coefficient is selected as the aerodynamic objective. Given the trend that higher lift coefficients often correspond to increased noise levels, a multi-objective optimization algorithm is employed to find a trade-off solution for these conflicting objectives. In this problem, the design parameters that define the geometry of the airfoil \( \Gamma \) are the control variables that are optimized under fixed operating conditions. The following objectives and constraints are selected:
\begin{align}
\begin{aligned}
 S^*= \text{argmin} & \ \left\{-C_L(S,\mathcal{Y}), OASPL(S,\mathcal{Y})\right\} \\
 \text{s. t.} \\
 & \ \frac{\partial{p}}{\partial{x}}(S,\mathcal{Y}) \geq 0 , \hspace{1.1cm} \\
 & \ S_{i}^{L} \leq S_{i} \leq S_{i}^{U}, \hspace{0.78 cm} i = 1,2,3.
\end{aligned}
\end{align}
where \( -C_L \) and \( OASPL \) are the aerodynamic and acoustic objective functions, \( S \) represents the set of design variables controlling the airfoil geometry profile \( \Gamma \), and \( \mathcal{Y} \) refers to the fixed flow conditions (e.g., Reynolds number and angle of attack). The geometric parameters controlling the airfoil deformation are \( M \) (maximum camber), \( P \) (position of maximum camber), and \( XX \) (maximum thickness), which define the NACA-4 digit series airfoils. The chord length is fixed for all deformed shapes. The optimal airfoil geometry, \( \Gamma^* \), is determined based on the optimized design variables \( S^* \). Once the geometric parameters are defined, airfoil shapes are modified by varying these parameters within the defined bounds (\( S_{i}^L, S_{i}^U \)). It should be noted that the upper and lower bounds for the geometric design variables are selected to ensure they fall within the range used to train our GNN model (Table \ref{tab:dataset}). Details on generating the airfoil profiles are provided in \ref{sec:Appendix-B}.

\begin{figure*}[h]
    \centering
    \includegraphics[width=1\linewidth]{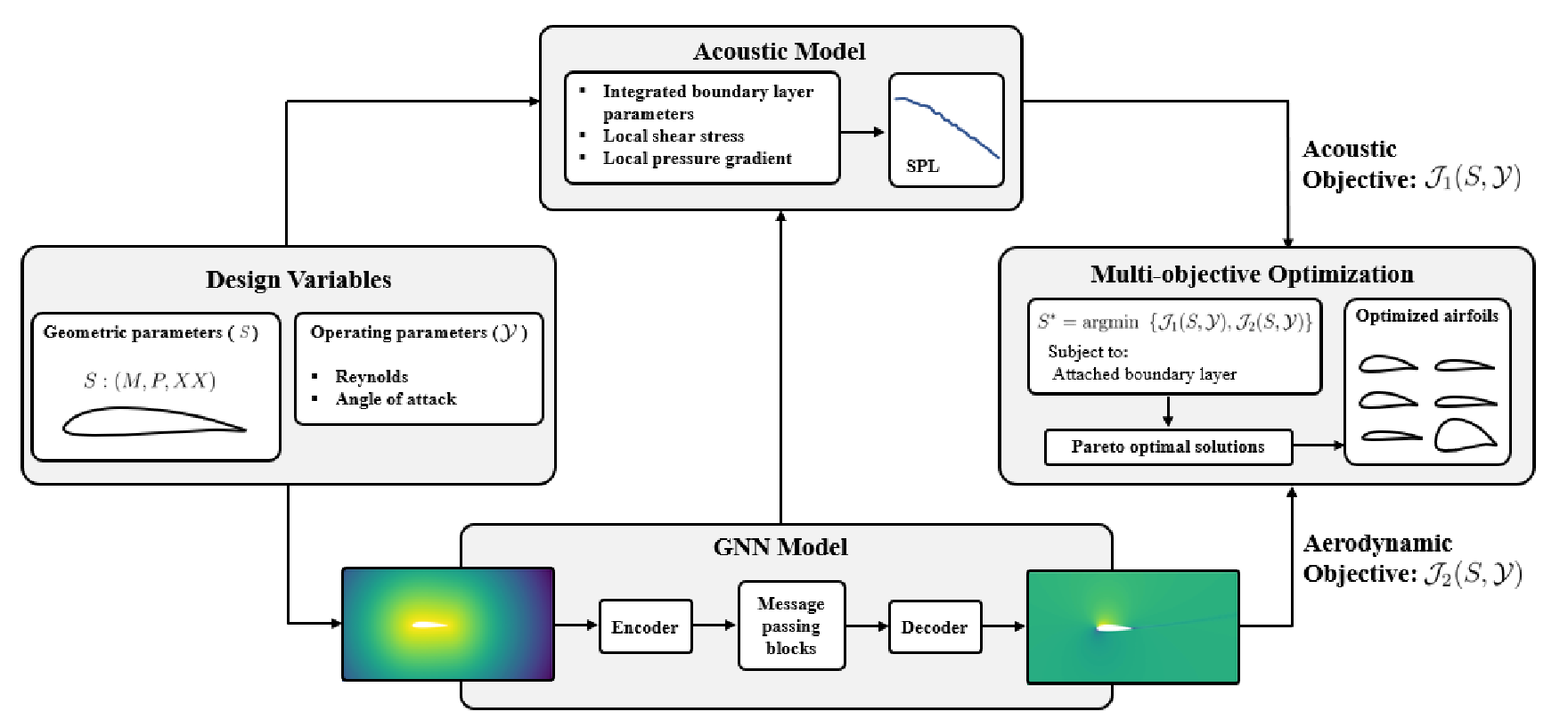}
    \captionsetup{font=small}
    \caption{Optimization framework.}
    \label{optim}
\end{figure*}

Figure \ref{optim} illustrates the optimization framework. For each combination of geometric and operating parameters, the trained GNN model predicts the full flow field. From these predictions, the aerodynamic objective function (\( -C_L \)) and the boundary layer parameters required to compute the power spectral density for \( OASPL \) are extracted. 
Since our acoustic model (Amiet’s theory) is valid only for attached boundary conditions, the optimization process enforces a positive pressure gradient (\( \frac{\partial p}{\partial x} \geq 0 \)) to avoid flow separation. Note that we do not impose any geometric constraints, as we generate airfoils using the NACA formulation within a reasonable range for geometric parameters. This ensures realistic airfoil geometries, where the upper and lower surfaces do not intersect, and the first half of the airfoil is thicker than the second half, providing a reasonable aerodynamic configuration.

To efficiently explore the global parameter space, we employ the evolutionary algorithm Non-dominated Sorting Genetic Algorithm II (NSGA-II) \cite{deb2000fast} for performing the multi-objective optimization. NSGA-II is a population-based evolutionary heuristic algorithm that follows the general framework of genetic algorithms with improvements in mating and survival selection mechanisms. Unlike single-objective optimization, which identifies a single best solution, NSGA-II preserves a diverse set of potential solutions along the Pareto front. A more detailed discussion of the NSGA-II implementation used in this study can be found in \cite{deb2000fast}.

Once the Pareto optimal set has been acquired, multiple criteria decision-making methods are applied to systematically evaluate and select a trade-off solution among the Pareto optimal candidates. One of the most commonly used multiple criteria decision-making techniques is the Technique for Order of Preference by Similarity to Ideal Solution (TOPSIS), which ranks solutions on the Pareto front based on their distances from the ideal and worst-case scenarios \cite{hwang1981methods}. Specifically, this method assesses solutions by measuring their proximity to the ideal solution and their distance from the negative ideal solution. In our study, the ideal solution is defined as the solution with the lowest noise intensity levels and the highest lift coefficient, while the negative ideal solution represents the solution with the maximum noise and minimum lift coefficient.
The TOPSIS algorithm is implemented in a MATLAB environment. To facilitate consistent calculation and processing, the initial data matrix is transformed into a forward matrix, followed by normalization to remove dimensional biases. Subsequently, the normalized data is analyzed to determine the positive and negative ideal solutions. The separation distances between each solution and both the ideal solution and the negative ideal solution are computed, allowing for the calculation of the relative closeness coefficient for each solution. The solutions are then ranked in descending order based on this coefficient, with the highest-ranked solution identified as the trade-off solution.

\section{Results and discussion}\label{sec:results_discussion}
In this section, we first demonstrate the capability of the designed GNN architecture to make data-driven predictions of flow fields for various airfoil profiles under different operating conditions. We then present results for aerodynamic performance statistics across the entire test dataset. Additionally, we showcase the network's ability to predict boundary layer characteristics and the resulting sound pressure levels in the far field.

\subsection{Flow field physical prediction}\label{sec:flow_field_prediction}
\begin{figure*}[h!]
\centering
\includegraphics[width=1\linewidth]{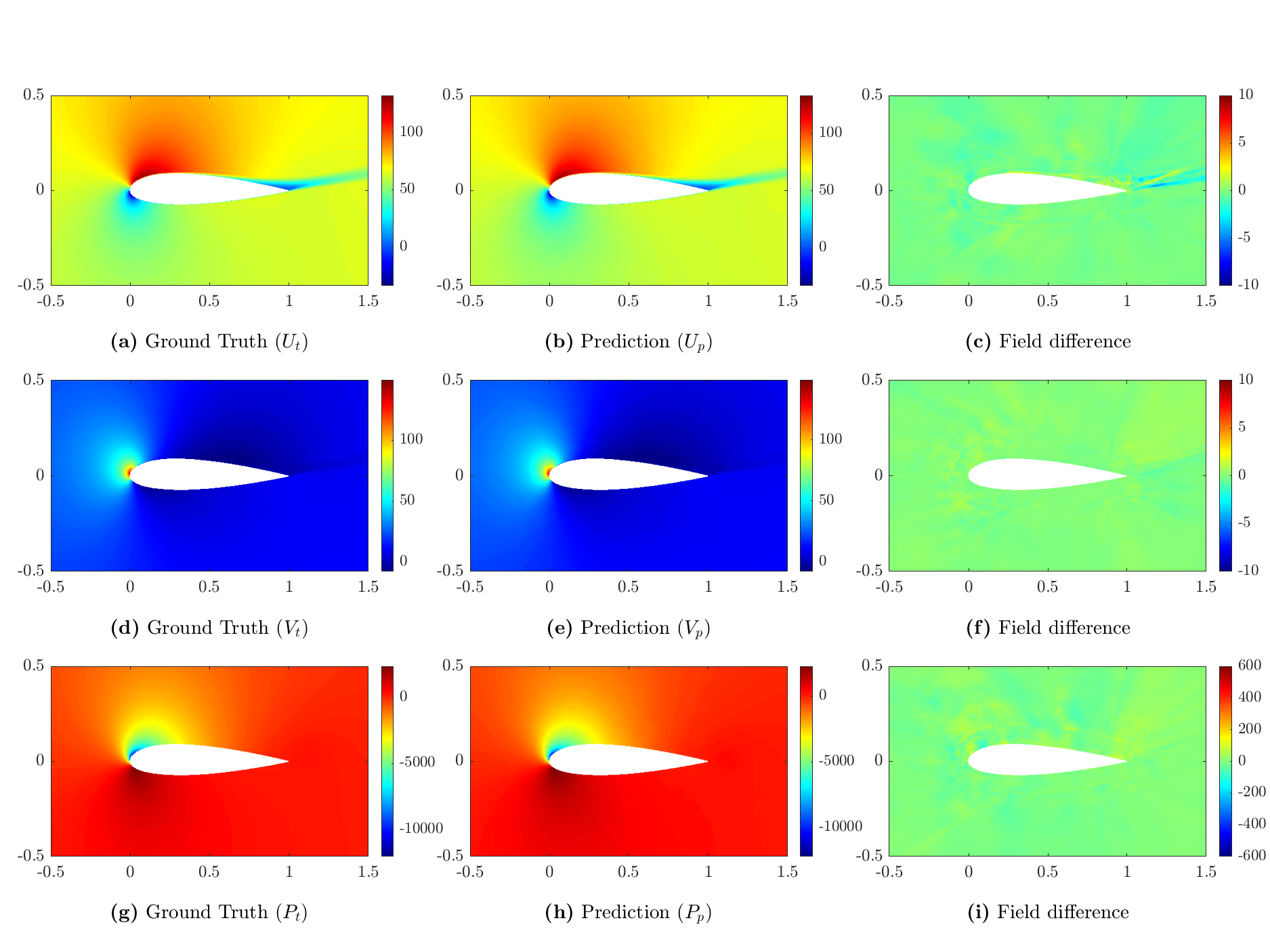}
\caption{\label{uvp_case_a} The pressure and velocity fields from the ground truth CFD simulations (left), the corresponding predictions from our proposed GNN model (middle), and their absolute difference (right) for Case I: $L = 0.899, P=4.541, S=1.0, XX=16.072, Re = 4.4222\times10^6, AOA=13.466$.}
\end{figure*}

\begin{figure*}[h!]
\centering
\includegraphics[width=1\linewidth]{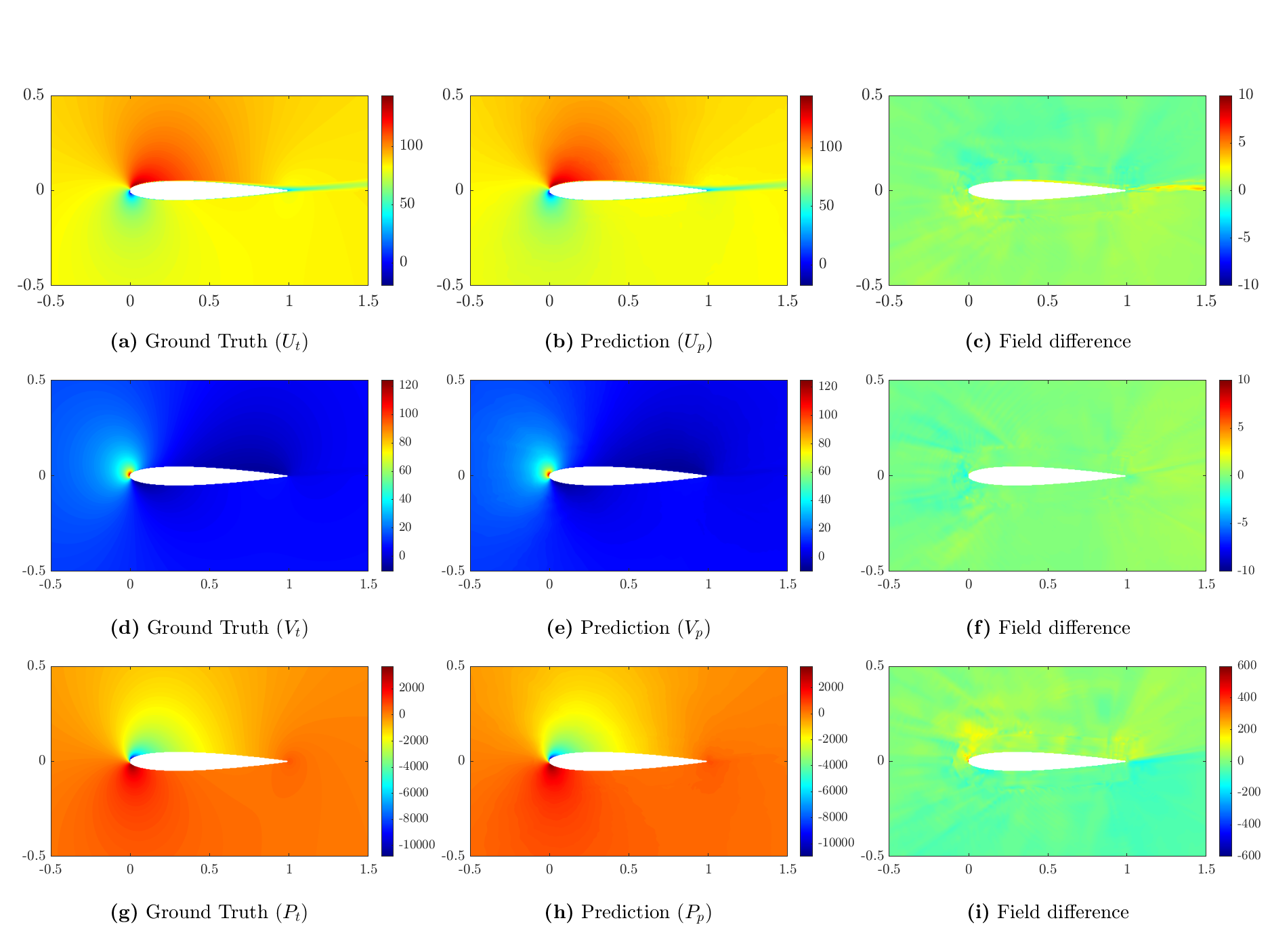}
\caption{\label{uvp_case_b} The pressure and velocity fields from the ground truth CFD simulations (left), the corresponding predictions from our proposed GNN model (middle), and their absolute difference (right) for Case II: $M = 2.499$, $P = 0.0$, $XX = 9.618$, $Re = 5.5779 \times 10^6$, $AOA = 6.764^\circ$.}
\end{figure*}

We evaluate the performance of the proposed GNN-surrogate model for shape optimization on 200 cases in the test dataset \( \left\{ \Gamma_{n}, \mathcal{Y}_n \right\}_{n=\left\{1, \dots,200\right\}} \), which are unseen by the GNN. 
Figures \ref{uvp_case_a} and \ref{uvp_case_b} illustrate the predicted and ground truth fields of the pressure and velocity components for two representative test cases, which we refer to as Case I and Case II, with high and moderate angles of attack, respectively. The GNN predictions show a close visual correspondence between the ground-truth CFD simulation results for both pressure and velocity. The model accurately captures key flow features, such as the local high-pressure region at the airfoil's leading edge, the high-velocity region along its upper surface, and the wake behind the trailing edge. Minor errors are observed in the x-component velocity predictions in the wake region: however, the predicted pressure fields align closely with the CFD simulations.
The absolute field difference, calculated as the pointwise difference between the predicted and ground truth values, highlights areas where the model may underestimate or overestimate certain flow characteristics. Despite these localized differences, the model demonstrates high predictive accuracy, particularly for pressure distribution and high-velocity regions. These results demonstrate the potential of the GNN-surrogate model as an efficient alternative to traditional CFD methods for evaluating airfoil performance with reasonable precision.

\begin{figure*}[h!]
\centering
\includegraphics[width=1\linewidth]{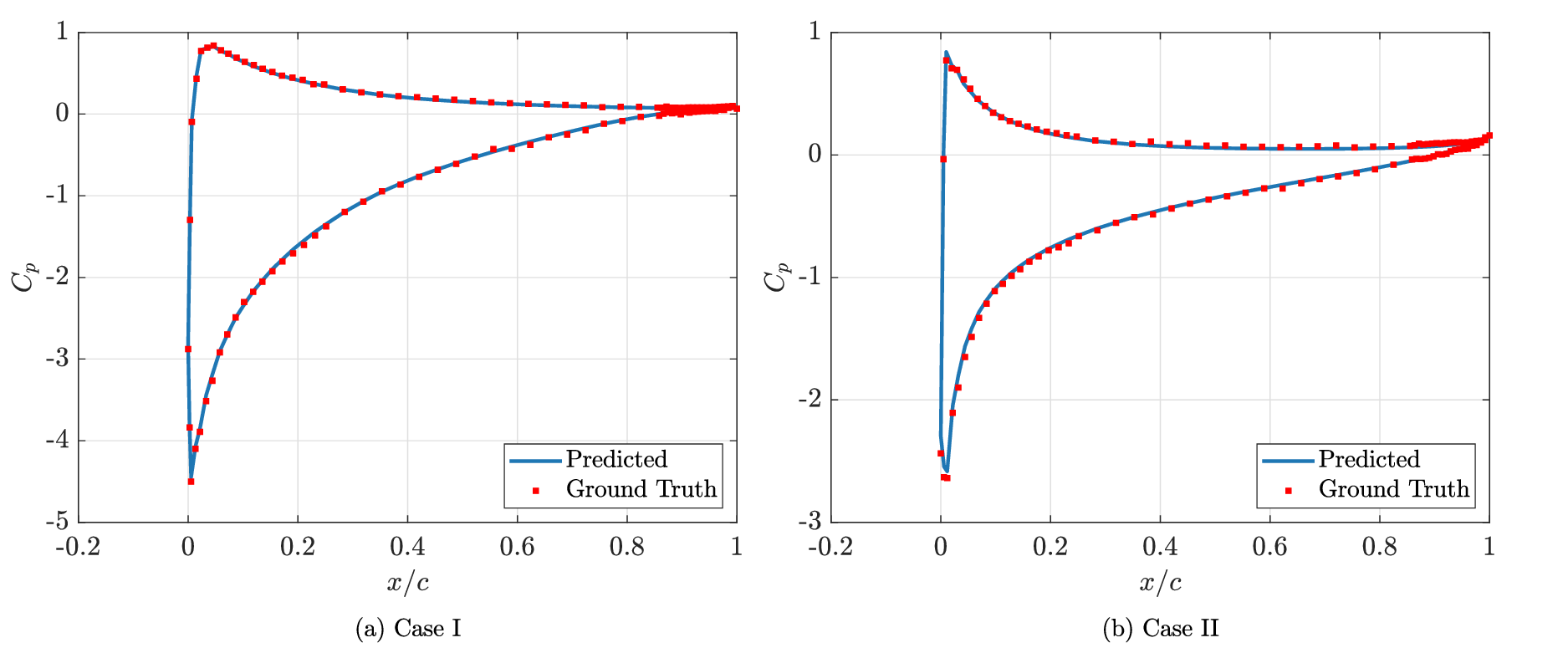}%
\caption{\label{cp} Comparison of the ground truth and predicted pressure distribution along the airfoil surface, where the predicted pressure coefficients are derived from the flow field predictions of the proposed GNN, for two test cases: (a) Case I; (b) Case II.}
\end{figure*}

To evaluate the aerodynamic performance of airfoils, the one-dimensional pressure distribution along the surface is a key factor. Using the proposed GNN framework, this distribution can be predicted with high accuracy and efficiency. Figure \ref{cp} compares the ground truth with the predicted pressure distribution across the airfoil surface for the two test cases, Case I and Case II, as described above. The predicted distributions closely match the true values obtained from the CFD simulations, with negligible error. 

\begin{figure*}[h!]
\centering
\includegraphics[width=1\linewidth]{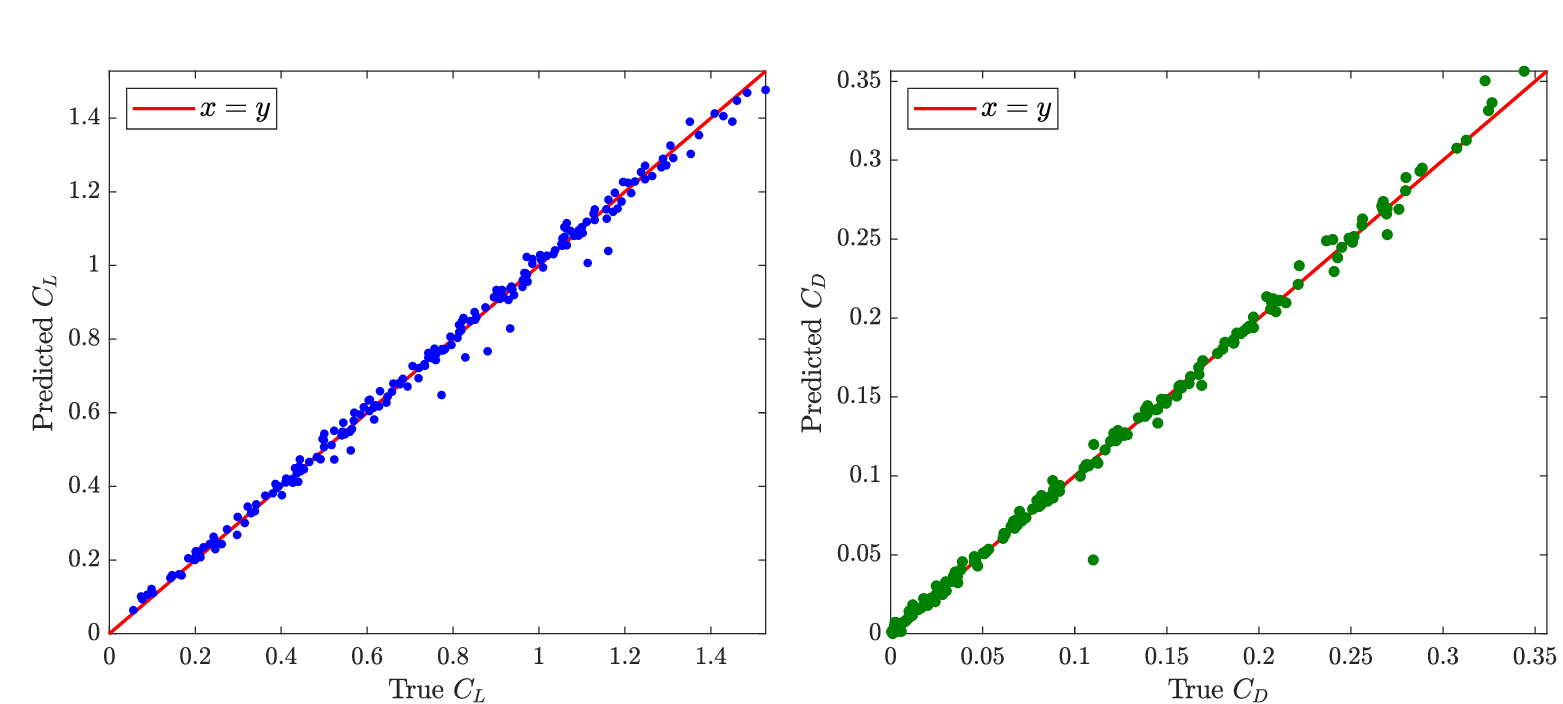}
\caption{\label{cl_cd} Cross-plots of the true versus predicted performance parameters for the test dataset; the 45-degree line indicates perfect agreement.}
\end{figure*}

Following this, we assess the performance of the GNN in terms of lift and drag coefficients, which are obtained from surface integrations over the airfoil. The cross plots for these cumulative quantities are presented in Figure \ref{cl_cd}. The 45-degree lines represent perfect agreement. The predictions for the lift coefficient demonstrate high accuracy, with most sample points aligning closely with the 45-degree line ($y = x$). Some scatter is evident in these plots and a few outliers appear in the drag coefficient plot; however, the overall level of agreement in both quantities is satisfactory. This observation is significant because aerodynamic performance computations used in optimization rely heavily on these quantities.

\begin{figure*}[h!]
\centering
\includegraphics[width=1\linewidth]{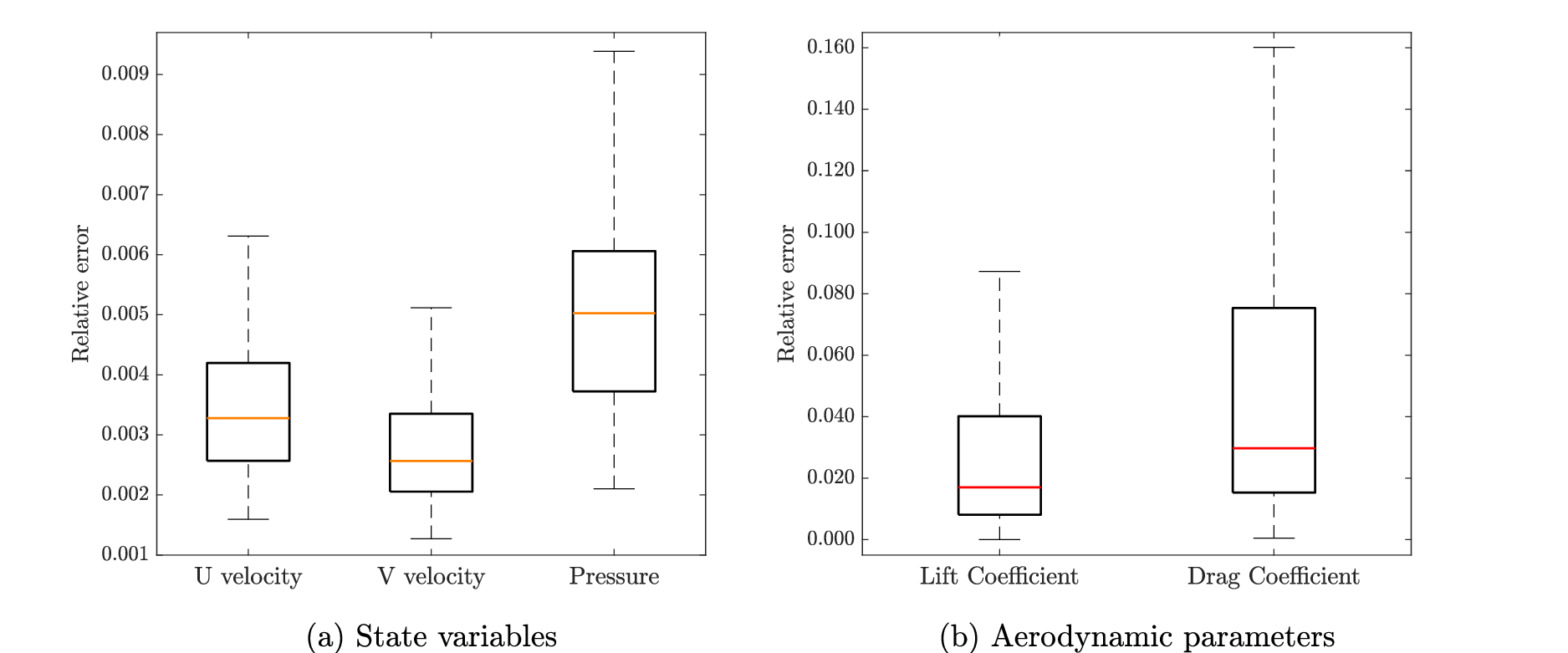}
\caption{Box plots of the relative errors for state variables (pressure and velocity components) and aerodynamic parameters (lift and drag coefficients) across 200 test cases. The boxes represent the \( P_{90} \), \( P_{75} \), \( P_{50} \), \( P_{25} \), and \( P_{10} \) percentiles. The error calculations are provided in Eqs. \ref{error:estate} and \ref{error:integ}.}
\label{error}
\end{figure*}

The relative errors for the 200 test cases, shown in the box plots of Figure \ref{error}, provide an overview of the error statistics for the state variables (pressure and velocity components) on the left and the performance parameters (lift and drag coefficients) on the right. In these box plots, the upper and lower whiskers represent the 90th  \( \left(P_{90}\right)\) and 10th \( \left(P_{10}\right)\) percentiles, while the upper and lower edges of the boxes correspond to the 75th \( \left(P_{75}\right)\) and 25th \( \left(P_{25}\right)\) percentiles, respectively. The median error \( \left(P_{50}\right)\) is indicated by the line within each box. As illustrated in Figure \ref{error} (a), the errors for the state variables are notably low, with \( \left(P_{90}\right)\) errors for pressure and saturation at approximately 0.5\%, 0.43\%, and 0.8\%. The relative errors for the lift and drag coefficients shown in Figure \ref{error} (b) are higher, ranging from 9\% to 16\%. The median errors for these coefficients are around 1.8\% and 3\%, respectively, indicating good performance. Furthermore, the error bars indicate that the mean relative error for the lift coefficient is 2.5\%, with the majority of sample points exhibiting errors below 3.29\%. The prediction accuracy for the drag coefficient is generally better among samples with lower drag values; however, several samples with higher drag coefficients show reduced accuracy, suggesting the presence of outliers. Overall, these findings indicate that the errors in the lift and drag coefficients are sufficiently small such that the proposed GNN can be used for optimization to reliably predict the flow field for new geometries represented by SDFs with high accuracy. 

\begin{figure*}[h!]
\centering
\includegraphics[width=1\linewidth]{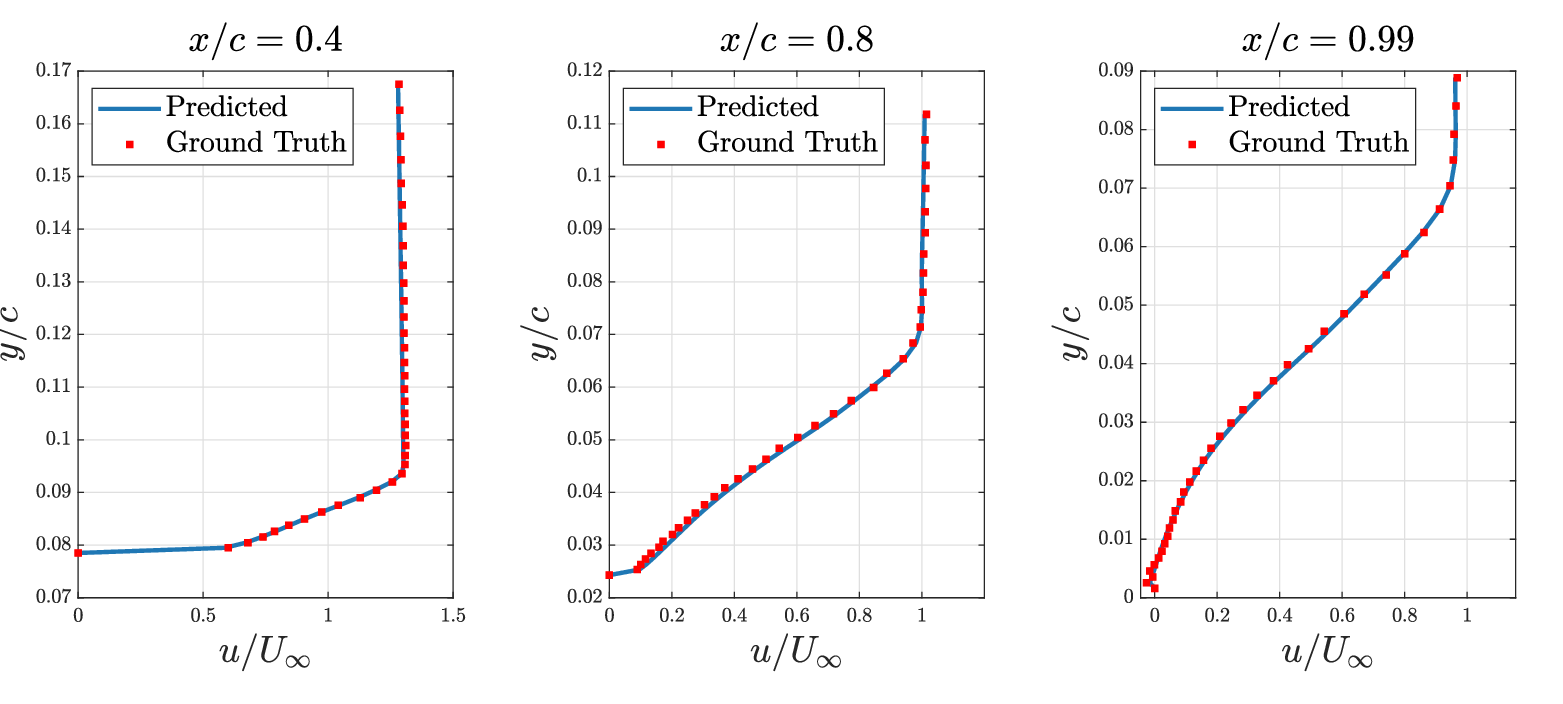}
\caption{\label{u_U_33} Comparison of true velocity profiles with velocity profiles extracted from flow field predictions obtained using our proposed GNN at different \(x/c \) positions for Case I, including flow separation at the trailing edge.}
\end{figure*}

\begin{figure*}[h!]
\centering
\includegraphics[width=1\linewidth]{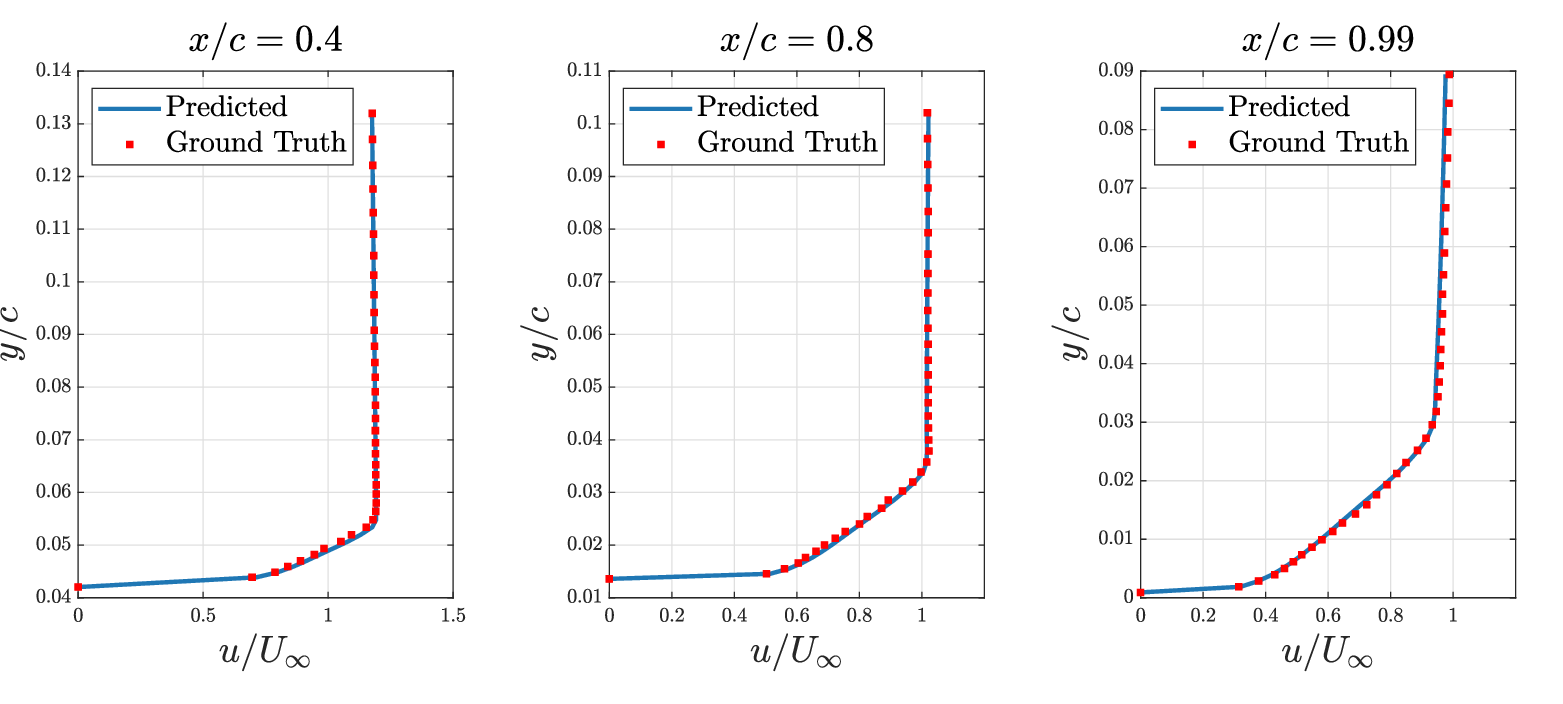}
\caption{\label{u_U_39} Comparison of true velocity profiles with velocity profiles extracted from flow field predictions obtained using our proposed GNN at different \(x/c \) positions for Case II.}
\end{figure*}

To highlight the effectiveness of our proposed GNN model in capturing the boundary layer region, Figures \ref{u_U_33} and \ref{u_U_39} illustrate the velocity profiles on the upper side of the airfoil for the above two cases Case I and Case II at various \( x/c \) positions (pressure and velocity fields for these cases are shown in Figures \ref{uvp_case_a} and \ref{uvp_case_b}). The predicted velocity profiles closely match the CFD results, accurately capturing sharp velocity variations near the airfoil surface. Additionally, the model can accurately capture separation effects, particularly in Case I, which involves a high angle of attack. Overall, the distribution characteristics of the velocity profiles at different \( x/c \) positions are predicted with high accuracy. With the boundary layer characteristics effectively captured, we present the results for the integrated boundary layer characteristics to predict acoustic levels in the following section.

\subsection{Integrated boundary layer characteristics and acoustic prediction}\label{sec:boundary_layer_acoustic}
\begin{figure*}[h!]
\centering
\includegraphics[width=1\linewidth]{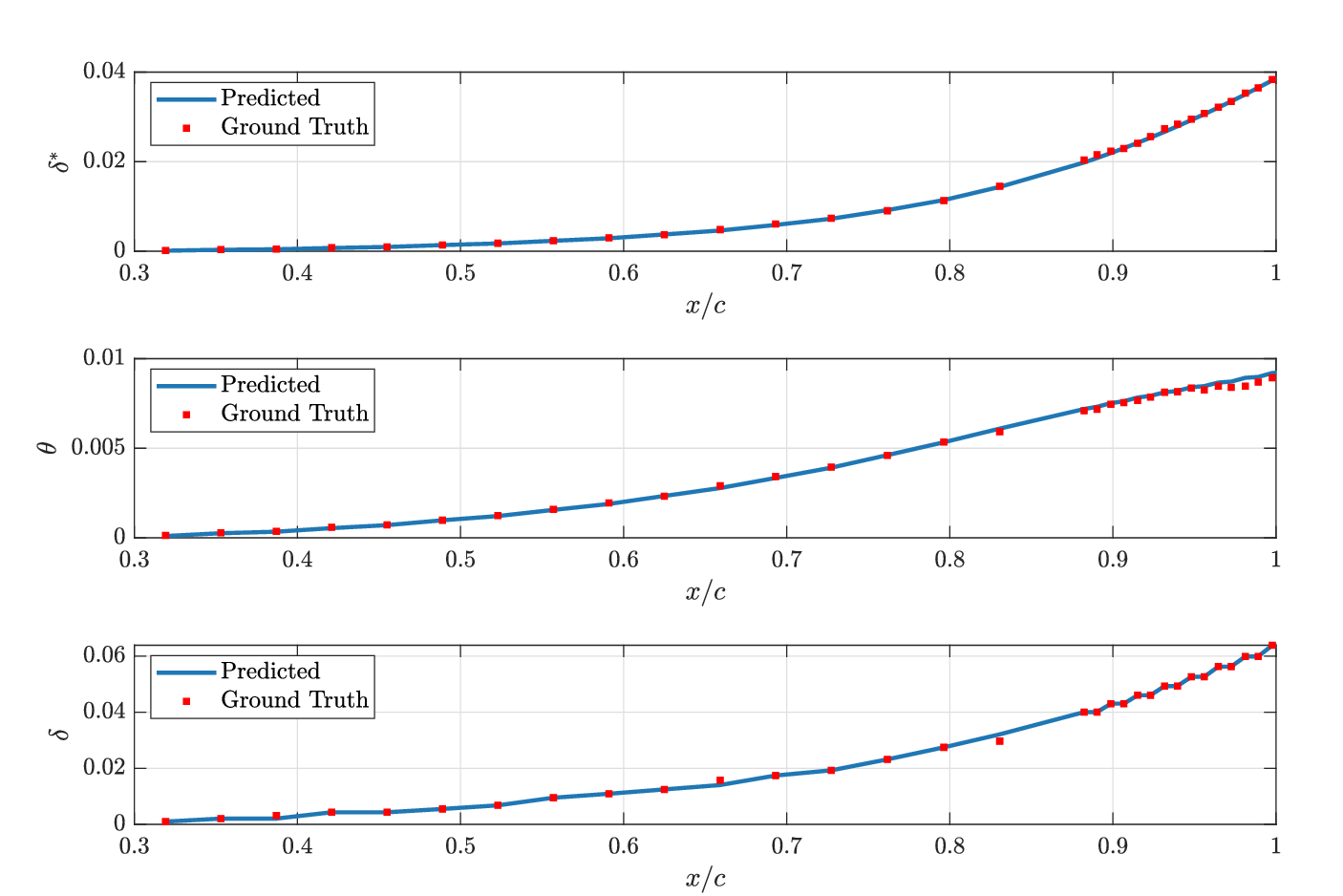}
\caption{\label{bl_33} Comparison of boundary layer thickness, momentum thickness, and displacement thickness obtained from ground truth CFD simulations and those extracted from flow field predicted by our proposed GNN model for Case I.}
\end{figure*}

\begin{figure*}[h!]
\centering
\includegraphics[width=1\linewidth]{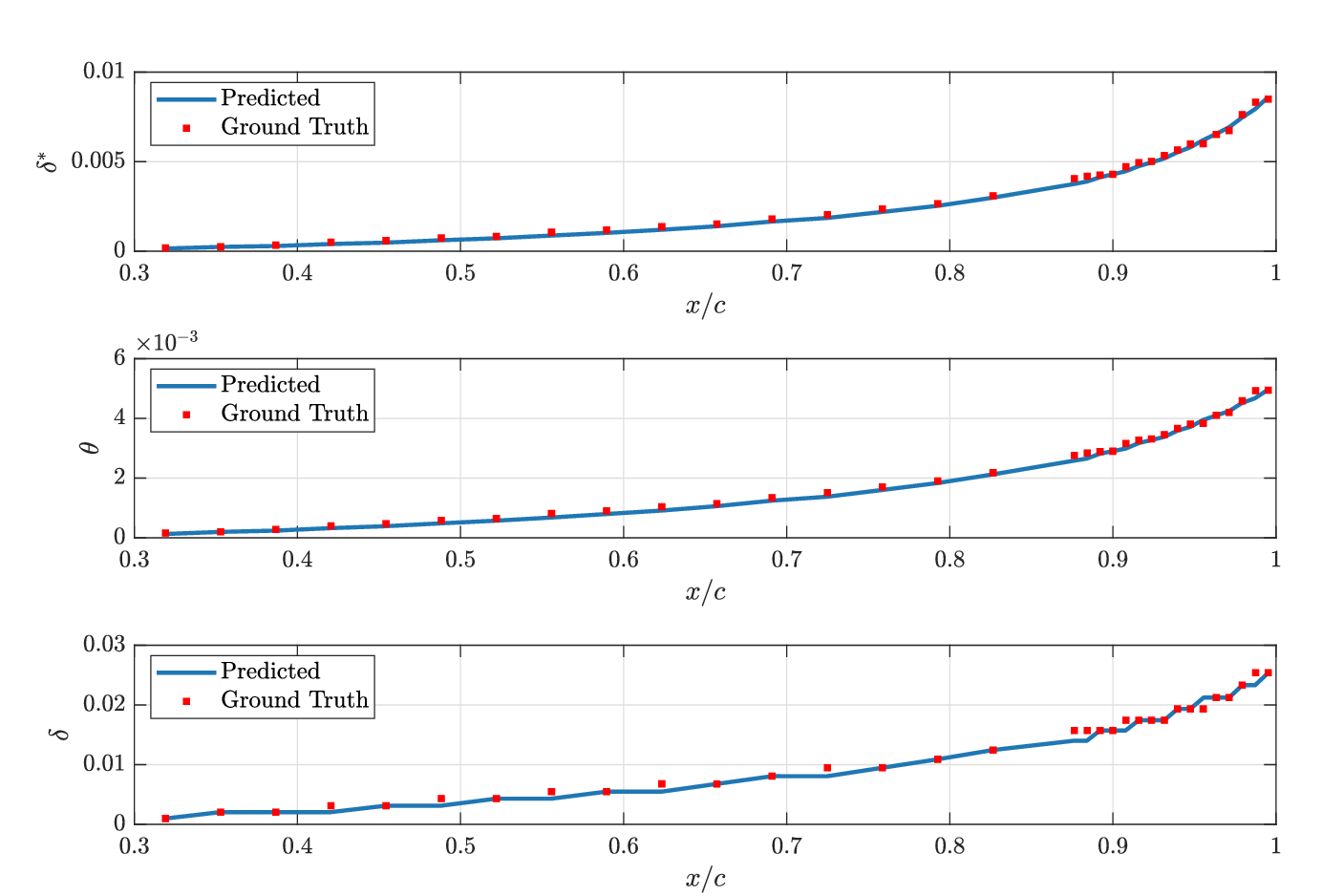}
\caption{\label{bl_39} Comparison of boundary layer thickness, momentum thickness, and displacement thickness obtained from ground truth CFD simulations and those extracted from flow field predicted by our proposed GNN model for Case II. These data are used to estimate the sound pressure level from Eq. \ref{spp_log}.}
\end{figure*}

The flow field prediction analysis demonstrates that our GNN-based surrogate model accurately captures the characteristics of the physical field, providing a solid foundation for the prediction of aerodynamic and acoustic performance. Using the predicted flow fields, we extract the boundary layer parameters necessary to estimate the terms in the power spectral density (Eq. \ref{spp}) for the test dataset \( \left\{ \Gamma_{n}, \mathcal{Y}_n \right\}_{n=\left\{1, \dots,200\right\}} \).  Figures \ref{bl_33} and \ref{bl_39} illustrate the characteristics of the integrated boundary layer required for the acoustic prediction for Cases I and II, showing that the boundary layer parameters \( \delta \), \( \delta^* \), and \( \theta \) are accurately captured and align well with the CFD results. 

\begin{figure*}[h!]
\centering
\includegraphics[width=1\linewidth]{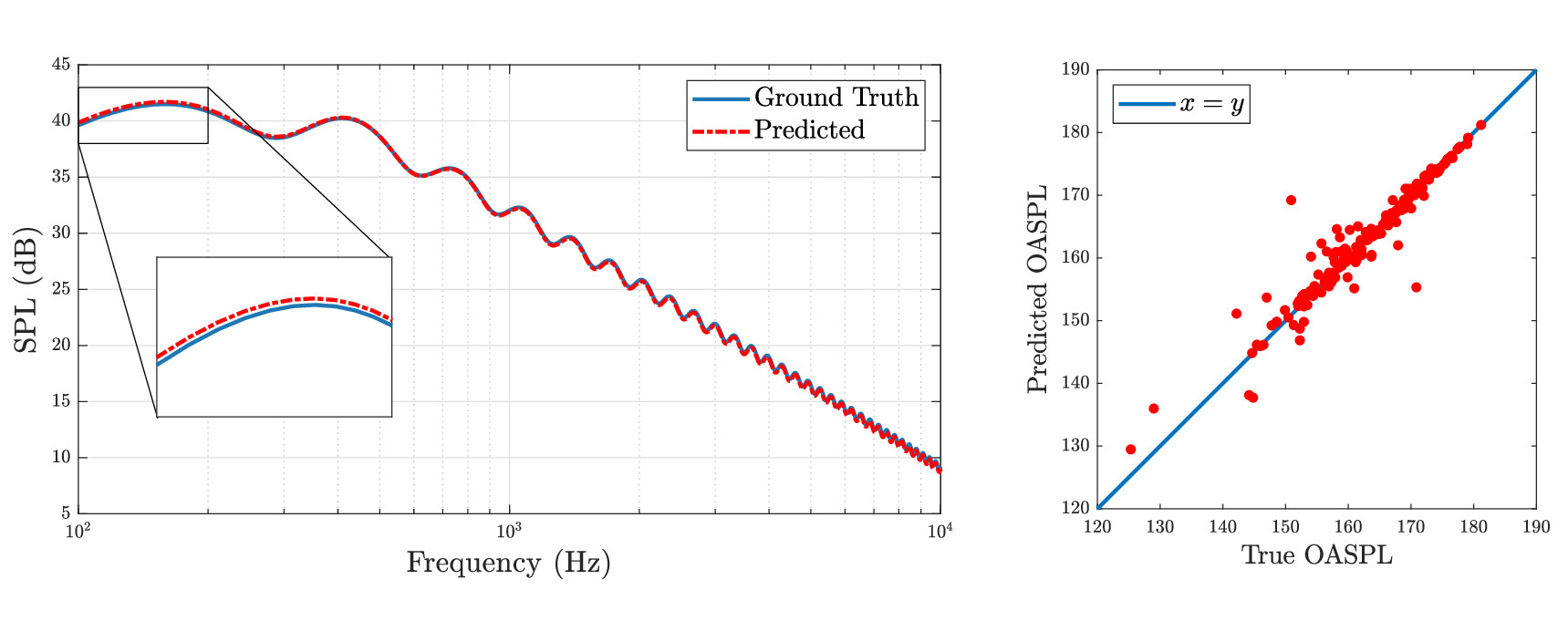}
\caption{\label{spl_b} The resulting sound pressure level from the integrated boundary layer parameters for Case II and the cross-plot of true versus predicted sound pressure levels for all cases with attached boundary layers in the test dataset.}
\end{figure*}

Since our acoustic model is valid only for attached boundary layers, all results in the remainder of this section are provided for cases with an attached boundary layer. As described in Section 4, the integrated boundary layer parameters are used to compute the resulting overall A-weighted sound pressure level. The integrated boundary layer parameters and the corresponding overall A-weighted sound pressure level are then compared with the values obtained from the CFD simulations. Table \ref{tab:error:oaspl} presents the mean relative errors for the integrated boundary layer parameters and the corresponding overall A-weighted sound pressure level for test cases with the attached boundary layer in the test dataset. As can be observed, the resulting errors are 8.32\%, 7.08\%, 8.55\%, and 2.28\% for \( \delta \), \(\delta^*\), \(\theta \), and the overall A-weighted sound pressure level, respectively. These results further confirm the reliability of our proposed GNN method for predicting acoustic levels. Figure \ref{spl_b} illustrates the sound pressure level for Case II, along with the 45-degree line for the A-weighted overall sound pressure level for all cases with an attached boundary layer in the test dataset. The figure demonstrates high accuracy, with most data points closely aligning with the 45-degree line (\( y = x \)). As observed, our proposed GNN shows excellent qualitative agreement with the sound pressure level obtained from the CFD simulations.

\begin{table}[h]
    \centering
    \captionsetup{font=small, justification=centering}
    \caption{Mean relative error of integrated boundary layer parameters and the resulting overall A-weighted sound pressure level for test cases with attached boundary layers.}
    \label{tab:error:oaspl}
    \small
    \setlength{\tabcolsep}{4pt}
    \renewcommand{\arraystretch}{1.3}
    \begin{tabular}{l c c c c}
        \toprule[1pt]
        Parameter & $\delta$ & $\delta^*$ & $\theta$ & OASPL \\
        \midrule
        Mean Relative Error & 8.32\% & 7.08\% & 8.55\% & 2.28\% \\
        \bottomrule[1pt]
    \end{tabular}
\end{table}

\subsection{Optimization Process} \label{subsec:optimization_process}
The \(x\)-\(y\) locations of the airfoil geometry, defined by three design variables corresponding to the \(M\), \(P\), and \(XX\) parameters, are optimized to enhance lift and noise performance using the proposed GNN-based prediction model. The initial population of 100 individuals is generated using the Latin Hypercube Sampling method \cite{mckay1979lhs} to ensure an evenly distributed sample across the design space. The optimization runs for a maximum of 100 generations with a mutation probability of 0.9. All optimizations are conducted using NSGA-II evolutionary algorithm implemented in the \texttt{pymoo} package \cite{blank2020pymoo}.
As detailed in Sections \ref{sec:boundary_layer_acoustic} and \ref{sec:flow_field_prediction}, the proposed GNN model provides reliable aerodynamic and acoustic predictions across a wide range of geometries. Therefore, the range of all three design parameters is set within the bounds used for GNN training. For this study, we use an operating condition of  \(\text{Re} = 2 \times 10^6 \) and  \(\alpha = 5^\circ\). Since our GNN model is trained for Reynolds numbers between 2 and 6 million and angles of attack between 5 and 15 degrees, any combination within these ranges can be chosen as the operating condition. 

\begin{table}[h!]
    \centering
    \captionsetup{font=small, justification=centering, skip=6pt}
    \caption{Design variables and objective function values for the three optimal cases obtained using the GNN-based fluid-acoustic shape optimization framework.}
    \label{tab:5}
    \small
    \setlength{\tabcolsep}{8pt}
\renewcommand{\arraystretch}{1.3} 
    \begin{tabular}{l c c c c c c}
        \toprule[1pt]
        &  \multicolumn{3}{c}{Design Variables} & & \multicolumn{2}{c}{Objective Functions} \\
        \cmidrule(lr){2-7} 
         & \( M \) & \(P\) & \(XX \)& & \(C_L\) & \(OASPL\) (dBA) \\
        \midrule
        Lift-dominated Solution A & 5 & 5 & 10 & & 1.3185 & 113.7901 \\
        Noise-dominated Solution B & 5 & 3.61 & 14.7 & & 1.2299 & 97.9690 \\
        Optimal Solution C & 5 & 4.94 & 14.71 & & 1.3111 & 99.9020 \\
        \bottomrule[1pt]
    \end{tabular}
\end{table}

\begin{figure*}[h!]
\centering
\includegraphics[width=0.6\linewidth]{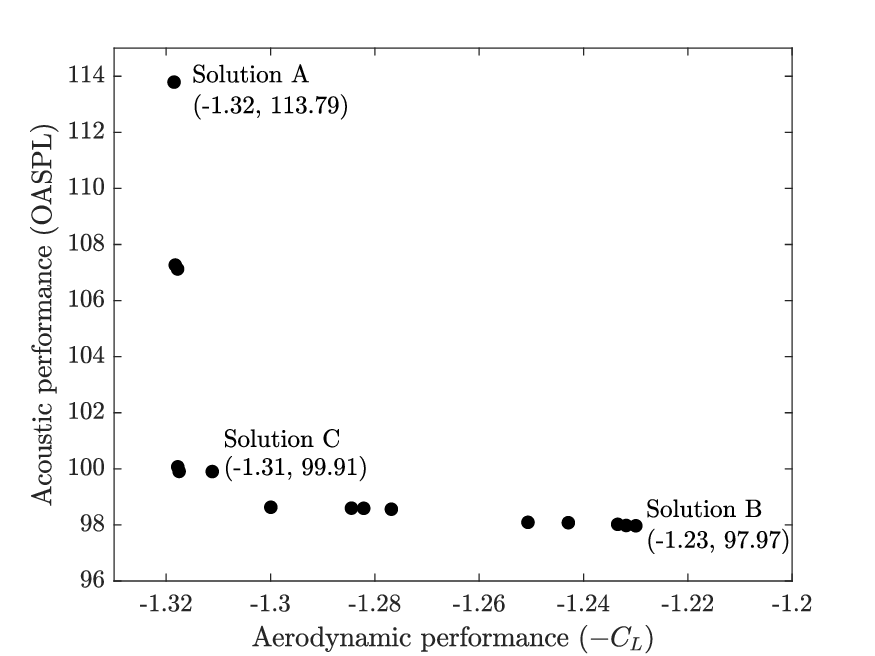}
\caption{\label{pareto} Pareto front optimal solution.}
\end{figure*}

Figure \ref{pareto} demonstrates the Pareto front, consisting of non-dominated optimal solutions obtained after 100 generations. Three particular solutions are highlighted as Solutions A , B, and C. Solution A corresponds to the lift-dominant objective function, while Solution B corresponds to the noise-dominant objective. Solution C, identified using the TOPSIS method, represents the optimal trade-off between the two objectives. It is evident that the overall sound pressure level gradually increases as the lift coefficient decreases. 

\begin{figure*}[h!]
\centering
\includegraphics[width=0.8\linewidth]{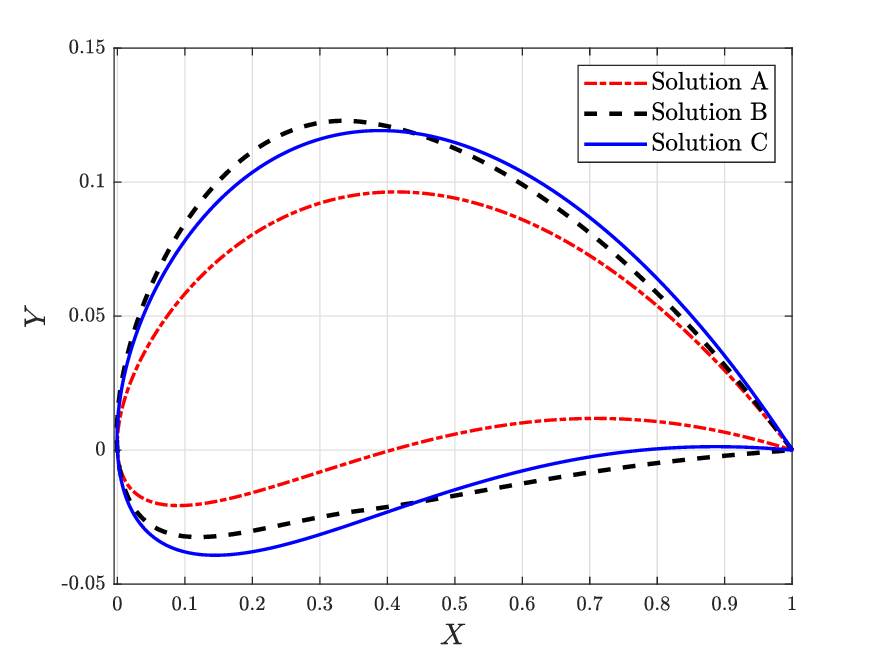}
\caption{\label{optimized_profile} Optimized airfoil geometries for Solutions A, B, and C, obtained using the proposed GNN-based shape optimization framework. The figure is enlarged to emphasize the differences between the geometries of the three solutions.}
\end{figure*}

The optimized airfoil shapes for Solutions A, B, and C on the Pareto front are illustrated in Figure \ref{optimized_profile}. Multiple optimization runs resulted in very similar final geometries. The optimized airfoils with higher aerodynamic performance exhibit a larger camber and are thinner compared to those with lower noise levels. This trend is expected, as the increase in camber creates a greater pressure difference between the upper and lower surfaces, enhancing lift. However, a higher camber can also lead to an increase in turbulence, contributing to higher noise levels generated by turbulent interactions. Solution A features a more rounded leading edge, which helps maintain smooth airflow over the airfoil, resulting in reduced noise. In contrast, solution B has sharper leading edges, which enhance lift but also increase noise due to the more turbulent flow. 

\begin{figure*}[h!]
\centering
\captionsetup{justification=centering}
\includegraphics[width=1\linewidth]{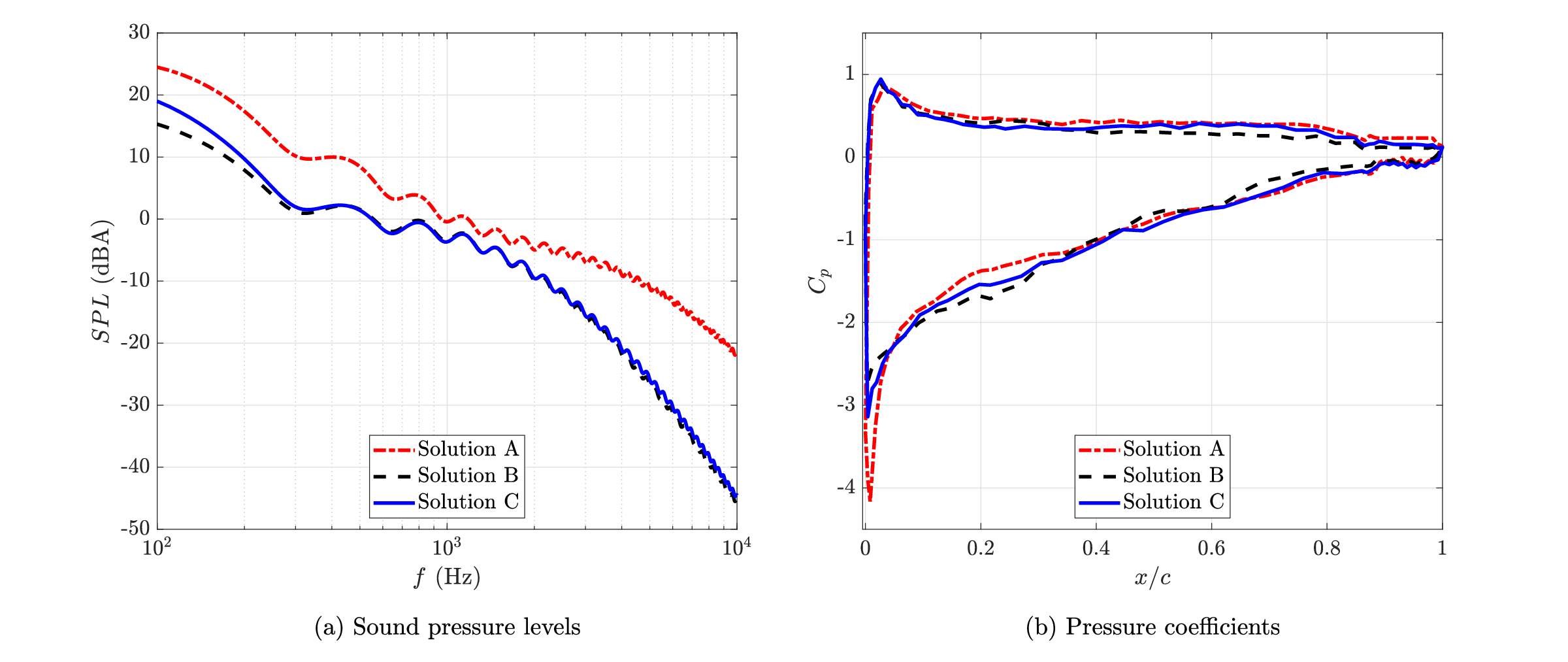}
\caption{\label{spl_cl_optimal} 
Comparison of aerodynamic and acoustic performance metrics, including sound pressure levels and lift coefficients, for three optimized solutions: A, B, and C. Solution A achieves the highest lift coefficient at the cost of maximum noise levels. Solution B minimizes noise but also results in the lowest lift coefficient, while Solution C represents a balanced compromise between aerodynamic efficiency and noise reduction.}
\end{figure*}

Figure \ref{spl_cl_optimal} illustrates the optimization results of sound pressure level and pressure coefficient values for the three optimal solutions:  lift-dominant Solution A, noise-dominant Solution B, and optimal Solution C. For each solution, the sound pressure level is computed following the methodology outlined in Section 3.1. As expected, the noise-dominant solution (Solution B) exhibits the lowest overall A-weighted sound pressure level of 97.97 dBA, with a minimum lift coefficient of 1.23. In contrast, the lift-dominant solution (Solution A) shows the highest overall A-weighted sound pressure level of 113.79 dBA, with a maximum lift coefficient of 1.32. This result demonstrates that, although the overall A-weighted sound pressure level is used as the acoustic objective function, noise is reduced across all frequency ranges. Solution C, representing the optimal trade-off, achieves a sound pressure level of 99.90 dBA and a lift coefficient of 1.31, offering a balance between acoustic and aerodynamic performance. It can be observed that as we move from the noise-dominant end (Solution B) to the lift-dominant end (Solution A) of the Pareto front, noise levels increase while the lift coefficient improves. The noise reduction from the noise dominant to the lift dominant solution is approximately 13.9\% (15.82 dBA), significantly reducing human annoyance levels, while the lift coefficient shows a 7.2\% increase from solution B to solution A. The values of all objective functions for the three optimal solutions are summarized in Table \ref{tab:5}. For completeness, the pressure coefficients of all configurations are compared in Figure \ref{spl_cl_optimal} (b). 

We now discuss the computational speed-up achieved by the proposed GNN in the context of neural network training, prediction, and optimization. The GNN training required approximately 40 hours on a single Nvidia A100 GPU. During optimization, evaluating the objective functions using the GNN, along with the necessary pre-and post-processing, takes about 3.75 seconds per candidate on the same GPU. Of this time, 2.27 seconds are spent generating the airfoil geometry based on the design variables and the signed distance function, including the extraction of the geometric features required as input for the GNN. The GNN prediction itself takes 0.45 seconds, while the acoustic prediction requires 0.29 seconds, and an additional 1.03 seconds is spent on further post-processing. The optimization procedure involved 10,000 simulations, resulting in an average total computation time of approximately 11 hours. In contrast, OpenFOAM simulations take approximately 25 minutes per simulation on 16 CPU cores of an AMD Ryzen\texttrademark\ Threadripper\texttrademark\ 3960X \cite{bonnet2022airfrans}. These results demonstrate that our GNN-based surrogate model can achieve a computational speed-up of three orders of magnitude compared to traditional CFD simulations. This significant advantage can substantially enhance the efficiency of the optimization process.

\section{Concluding remarks} \label{sec:concluding_remarks}
In this study, we proposed a novel data-driven surrogate method using a graph neural network for fluid-acoustic shape optimization applications. Our GNN-based surrogate method can handle arbitrary unstructured meshes without requiring modifications to the model. Using our approach, the network efficiently predicts the quantities of interest for a wide range of shapes based on the provided mesh and physical properties. Our goal has been to employ the proposed GNN-based surrogate model to alleviate the computational costs inherent in shape optimization and accurately predict flow quantities while capturing boundary layer characteristics. The trained model was evaluated on various geometries, Reynolds numbers, and angles of attack for 2D airfoils wherein the airfoil geometries were represented using signed distance functions. For a set of 300 test cases, the proposed model achieved median errors of approximately 0.5--1\% for pressure and velocity fields. The errors were slightly higher for the aerodynamic performance metrics, with 7\% for the lift coefficient and 4\% for the drag coefficient. To demonstrate the model's capability to accurately capture boundary layer phenomena, we presented velocity profiles at three locations along the airfoil surface, showcasing highly accurate predictions. While the model effectively captures local flow dynamics in the boundary layer, the predictive accuracy for the overall A-weighted sound pressure level is 2.3\%.

We demonstrated the efficacy of our optimization framework for fluid-acoustic airfoil shape optimization. The goal was to determine the optimal airfoil surface shapes under specific operating conditions, aiming to reduce trailing edge noise produced by an attached turbulent boundary layer while maximizing aerodynamic performance. We integrated our trained GNN model with the non-dominated sorting genetic algorithm to enable an efficient global search.  To identify a trade-off solution among the Pareto front solutions, we employed a multi-criteria decision-making method and three optimal geometry profiles, namely noise dominant, lift dominant, and a trade-off profile, were analyzed. Using the optimization framework, we found that the noise-dominant solution exhibited a 13.9\% reduction (15.82 dBA) in the overall A-weighted sound pressure level, while the lift-dominant solution showed a 7.2\% increase in lift coefficient. 

This study presents a promising methodology where GNNs efficiently and accurately establish nonlinear mappings between geometric design parameters, physical properties, and aerodynamic/hydrodynamic and acoustic performance metrics, enabling rapid shape optimization. The research represents a comprehensive and interdisciplinary effort that holds substantial scientific significance for advancing intelligent approaches in fluid mechanics and acoustics. The non-intrusive modeling and the implicit shape representation approach adopted here make this method applicable for a wide range of applications, including complex three-dimensional shapes. Furthermore, the GNN model makes this approach potentially scalable for larger, more complex problems. Such studies will be explored in future work to address shape optimization challenges in the aerospace and marine sectors.


\section*{Acknowledgement} \label{sec:acknowledgement}
The authors would like to acknowledge the Natural Sciences and Engineering Research Council of Canada (NSERC) and Seaspan Shipyards for the funding. This research was supported in part through computational resources and services provided by Advanced Research Computing at the University of British Columbia.

\appendix
\setcounter{equation}{0} 
\setcounter{figure}{0}

\section{Parameters in Amiet’s theory for aeroacoustics}\label{sec:Appendix-A}
\subsection{Aeroacoustic transfer function}
The term \( \mathscr{L} \left( \frac{\omega}{\overline{U}_{c}}, k_z = 0, x, y, U_{\infty}, \overline{U}_{c} \right)
\) is  the aeroacoustic transfer function that represents the relationship between wall pressure fluctuations caused by the impinging turbulence and the acoustic waves in the far field Ref \cite{roger2005back}, which is defined as:

\begin{equation}
\begin{aligned} \label{L}
|\mathscr{L}| = \frac{1}{\Theta}\Bigg| \left(1+i\right) \left\{ \sqrt{\frac{\frac{\overline{k}_{x}}{\mu}+M + 1}{1+\frac{x}{\sigma}}} E^*\left(2\mu \left( 1+ \frac{x}{\sigma} \right)\right)e^{-2i\Theta} - E^* \left( 2\left(\overline{k}_{x} + \mu (M+1)\right)\right) \right\} + 1 \Bigg|,
\end{aligned}  
\end{equation}
where 
\begin{equation}
\begin{aligned} \label{Theta_big}
\Theta \equiv \overline{k}_{x} + \mu \left( M - \frac{x}{\sigma} \right),
\end{aligned}  
\end{equation}
Here \( \mathrm{E}^*\left( x \right)
 \) represents the complex error function, which is a combination of Frensel integrals \( \mathrm{C}_2\left( x \right)
 \) and \( \mathrm{S}_2\left( x \right)
 \) \cite{abramowitz1965handbook}.
\begin{equation}
\begin{aligned} \label{E_star}
\mathrm{E}\left( x \right) &\equiv \frac{1}{\sqrt{2\pi}}\int_{0}^{x} \frac{e^{it}}{\sqrt{t}} \, dt = \mathrm{C}_2\left( x \right) - i \, \mathrm{S}_2\left( x \right).
\end{aligned}  
\end{equation}

\subsection{Spanwise correlation length}
\( \Lambda_{\mathrm{z}|\mathrm{pp}}
 \) is the spanwise correlation length, which is a measure of the distance in the spanwise direction where there is a high level of coherence in turbulent structures. This correlation length can be estimated from coherence measurements between surface pressures at multiple spanwise points. It is typically calculated using Corcos' model \cite{corcos1964structure}, where \( b_c = 1.47 \) is an empirical constant.
\begin{equation}
\begin{aligned} \label{lambda}
\Lambda_{z|pp} = b_{c} \frac{\overline{U}_c}{\omega}.
\end{aligned}
\end{equation}

\subsection{Surface pressure frequency spectrum}
The surface pressure frequency spectrum can be obtained from the following equation: 
\begin{equation}
\begin{aligned}
\Pi_{\mathrm{pp}} \left( k_{x}, k_{z}, \omega \right) &= \frac{4\pi \rho^2}{\Lambda_{z|pp} \left( \omega \right)} \int_{0}^{\delta} \Lambda_{y|vv} \left( y \right) \mathrm{U}_{c} \left( y \right) \left[ \frac{\partial \mathrm{U} \left( y \right)}{\partial y} \right]^2 \frac{\overline{u}^2_{y} \left( y \right)}{\mathrm{U}_{c}^2 \left( y \right)} \phi_{vv} \left( \frac{\omega}{\mathrm{U}_{c} \left( y \right)}, k_{z} = 0 \right) e^{-2 |k| y} \, dy,
\end{aligned}
\label{eq:pp_correlation}
\end{equation}
where \( \Lambda_{\mathrm{y}|\mathrm{vv}}
 \) is the transverse correlation of the normal velocity component. This integral length scale can be calculated using Schlichting's formula \cite{schlichting2016boundary} for the mixing length scale \( l_{\text{mix}}
 \) as follows:

\begin{equation}
\begin{aligned} \label{spanwise_correlation_length}
\Lambda_{y|vv} = \frac{l_{\text{mix}}}{K},
\end{aligned}
\end{equation}

\begin{align}
l_{\text{mix}} &= \frac{0.085 \delta \, \tanh \left( \frac{K}{0.085} \left( \frac{y}{\delta} \right) \right)}{\sqrt{1 + B \left( \frac{y}{\delta} \right)^6}},
\end{align}
in which \( K=0.38 \) is von-Karman constant and $B = 5$ is the level constant. \( \delta \) is the boudary layer tickness and can be calculated from the following emprical relation proposed by Drela \cite{drela1987viscous}:
\begin{equation}
\begin{aligned} \label{delta}
\delta &= \theta \left( 3.15 + \frac{1.75}{H - 1} \right) + \delta^*,
\end{aligned}
\end{equation}
where \( \delta^* \) is the boundary layer displacement thickness and \( \theta \) is the momentum thickness, both of which are derived from the velocity field generated by the GNN-based surrogate model. The ratio \( H = \delta^* / \theta \) is the kinematic shape factor.

Here $U_c$ represents the local convection velocity across the boundary layer, which is taken as \( U_c = 0.6 U_{\infty} \). The mean streamwise velocity outside the viscous-dominated near-wall region is the sum of a logarithmic component and a wake component, as modeled by von-Kármán. The variation of the normalized mean velocity 
\(
\mathrm{U}^{+} = \frac{\mathrm{U}(y)}{u_{\tau}},
\)
with the normalized distance from the wall 
\(
y^+ = \frac{y \, \mathrm{u_\tau}}{\nu},
\)
is therefore described by:
\begin{equation}
\mathrm{U}^+ \left( y \right) = \frac{1}{K} \ln \left( y^+ \right) + B + \frac{2 \Pi_{\mathrm{w}}}{K} \sin^2 \left( \frac{\pi y}{2 \delta} \right).
\label{eq:U+}
\end{equation}
where $u_\tau$ is the friction velocity, given by $ u_{\tau} = \sqrt{\tau_{w}/\rho} $, and $\tau_{w}$ is the local shear stress, which is calculated based on the velocity field generated by the GNN model. The last term in Eq. \ref{eq:U+} is the Coles wake factor $\Pi_w$ \cite{coles1956law}, which shows the effect of the adverse pressure gradient on the mean streamwise velocity $U(y)$ and is a function of the local pressure gradient $dC_p/dx$ and the wall shear stress $\tau_w$.
\begin{equation}
\begin{aligned} \label{pi}
\Pi_{w} = 0.8 \left( \frac{\delta^*}{\tau_{w}} \frac{dC_{p}}{dx} + 0.5 \right)^{3/4},
\end{aligned}
\end{equation}

The variation of the mean square turbulence velocity $\frac{u_{y}}{U(y)}$ is calculated based on the model proposed by Alfredsson \cite{alfredsson2011new}.

\begin{equation}
\frac{u_{x}(y)}{U(y)} = \left( a + b \left( \frac{U(y)}{U_{\infty}} \right) \right) Q \left( \frac{U(x)}{U_{\infty}} \right),
\end{equation}
where $a = 0.2909$ and $b = -0.2598$ are empirical constants. Since the mean streamwise $U(y)$ velocity approaches the free-stream velocty much faster than $u_{x}(y)$ in the free-stream region, the correction factor $Q$ is introduced as follows: 
\begin{equation}
Q \left( \frac{U(x)}{U_{\infty}} \right) = 1 - e^{-\gamma \left( 1 - \frac{U(y)}{U_{\infty}} \right)}.
\end{equation}
where $\gamma = 64$. The turbulence intensity in the normal directions is calculated as follows: 
\begin{equation}
\begin{aligned} \label{u2}
\overline{u^2_{y}} = \beta_{y}\overline{ u^2_{x}},
\end{aligned}
\end{equation}
where $\beta_{y} = 1/2$.
In Eq. \ref{eq:pp_correlation}, $\phi_{vv}$ is normalized vertical velocity spectrum and calculated as follows: 

\begin{equation}
\begin{aligned} \label{phi_vv}
\phi_{vv} = \frac{4}{9\pi}\frac{\beta_{x}\beta_{z}}{k^2_{e}}\frac{(\beta_x k_{x}/k_{e})^2+(\beta_z k_{z}/k_{e})^2}{[1+(\beta_x k_{x}/k_{e})^2+(\beta_z k_{z}/k_{e})^2]^{7/3}},
\end{aligned}
\end{equation}
where $\beta_{x} = 1$, $\beta_{z} = 3/4$ are streamwise and spanwise anisotropic turbulence factors, respectively. $k_{e}$ is the wavenumber of eddies that contains most of the energy and is inversly proportional to the integral legth scale as follows:
\begin{equation}
\begin{aligned} \label{ke}
k_{e}(y) = \frac{\sqrt{\pi}}{\Lambda_{x|uu}}\frac{\Gamma(5/6)}{\Gamma(1/3)},
\end{aligned}
\end{equation}
where $\Lambda_{x|uu} =2 \Lambda_{y|vv}$. For airfoils with large span the turbulence wavenumber $k_{z}=0$ and the one dimensional velocity spectra is modeled as follows; 
\begin{equation}
\begin{aligned} \label{phi_uu}
\phi_{uu} = \frac{\Gamma(5/6)}{\sqrt(\pi)\Gamma(1/3)}\frac{\beta_{1}}{k_{e}}\frac{1}{(1+(\beta_{x}k_{x}/k_{e})^2)^{5/6}}.
\end{aligned}
\end{equation}

\section{Details on Generating NACA Airfoils}\label{sec:Appendix-B}
The NACA four-digit series define the profile by a sequence MPXX, where M is the maximum camber as a percentage of the chord, P is the position of this maximum from the airfoil leading edge in tenths of the chord, and XX is the maximum thickness of the airfoil as a percentage of the chord. For example, the NACA 2412 airfoil has a maximum camber of 2\% located 40\% (0.4 chords) from the leading edge with a maximum thickness of 12\% of the chord. The leading edge of each airfoil is consistently positioned at the coordinates (0, 0), while the trailing edge is consistently located at (chord, 0) within the x-y plane. For the five digits series, each airfoil is defined by a sequence of LPSTT, where L is a single digit representing the theoretical optimal lift coefficient at the ideal angle of attack CLI = 0.15 L, and P is the position of maximum camber divided by 20, S is a boolean that indicating whether the camber is simple (S = 0) or reflex (S = 1). The last two digits of TT have the same definition as in the four digits case. For example, the NACA 12018 airfoil would give an airfoil with a maximum thickness of 18\% chord, maximum camber located at 10\% chord, with a lift coefficient of 0.15. Therefore, each four and 5-digit airfoil is generated based on 3 or 4 parameters. 

In both NACA four- and five-digit series, the thickness distribution is:

\begin{equation} \label{thickness}
T\left(x\right) = 5\tau c \left[ 
    0.2969 \sqrt{\left(\frac{x}{c}\right)} 
    - 0.126 \left(\frac{x}{c}\right) 
    - 0.3516 \left(\frac{x}{c}\right)^2 
    + 0.2843 \left(\frac{x}{c}\right)^3 
    - 0.1015 \left(\frac{x}{c}\right)^4 
\right].
\end{equation}
where \( c\) is the airfoil chord, \(x\) is the distance along the chord from the leading edge, \(\tau\) is the maximum thickness as a fraction of the chord defined by the two last digits.
\par
The camber line of a four-digit airfoil consists of two parabolas joined at the maximum camber location as follows:

\begin{equation}
y_{c}\left(x\right) = 
\begin{cases}
	\frac{m}{p^2} \left(\left(\frac{x}{c}\right) \left(2p - \left(\frac{x}{c}\right)\right)\right), & 0 \leq \left(\frac{x}{c}\right) \leq p, \\[10pt]
	\frac{m}{(1-p)^2} \left[\left(1 - 2p\right) + 2p\left(\frac{x}{c}\right) - \left(\frac{x}{c}\right)^2\right], & p < \left(\frac{x}{c}\right) \leq 1.
\end{cases}
\end{equation}
where \( m=\frac{M}{100} \) and \( p = \frac{P}{10} \). 
The equation for the camber line of a five-digit series is also split into two sections like the four-digit series, but the division between the two sections is not at the point of maximum camber. This function for the simple case (S = 0) is defined as follows:

\begin{equation}
y_{c}\left(x\right) = 
\begin{cases}
	\frac{k_1}{6}\left[\left(\frac{x}{c} \right)^{3} - 3m\left(\frac{x}{c} \right)^{2} + m^{2}\left(3-m\right)\left(\frac{x}{c} \right)\right], & \hspace{1cm} 0 \leq \left(\frac{x}{c} \right) \leq p \\[10pt]
    \frac{k_1 m^{3}}{6}\left(1 - \left(\frac{x}{c} \right)\right), & \hspace{1cm} p < \left(\frac{x}{c} \right) \leq 1
\end{cases}
\end{equation}

The camber line function for the reflex case (S = 1) is defined as follows:

\begin{equation}
y_{c}\left(x\right) = 
\begin{cases}
	\frac{k_1}{6}\left[\left(\left(\frac{x}{c}\right) - m\right)^{3} - \frac{k_{2}}{k_{1}}\left(1 - m\right)^{3}\left(\frac{x}{c}\right) - m^{3}\left(\frac{x}{c}\right) + m^{3}\right], & \hspace{1cm} 0 \leq \left(\frac{x}{c}\right) \leq p \\[10pt]
    \frac{k_1}{6}\left[\left(\frac{k_2}{k_1}\left(\frac{x}{c}\right) - m\right)^{3} - \frac{k_{2}}{k_{1}}\left(1 - m\right)^{3}\left(\frac{x}{c}\right) - m^{3}\left(\frac{x}{c}\right) + m^{3}\right], & \hspace{1cm} p < \left(\frac{x}{c}\right) \leq 1
\end{cases}
\end{equation}
where \( m \), \( k_{1}\), and \(k_{2}\) can obtained from the following:
\begin{equation}
\begin{aligned} \label{p}
	p = m (1-\sqrt{\frac{m}{3}}),
\end{aligned}
\end{equation}

\begin{equation}
\begin{aligned} \label{k1}
	k_{1} = \frac {C_{L}}{\frac{3m - 7m^{2} + 8m^{3} - 4m^{4}}{\sqrt{m\left(1 - m\right)}} - \frac{3}{2}\left(1 - 2m\right) \left(\frac{\pi}{2} - \mathrm{\arcsin}\left(1 - 2m\right)\right)}, 
\end{aligned}
\end{equation}

\begin{equation}
\begin{aligned} \label{k2}
	k_{2} = \frac{3\left(m - p\right)^{2} - m^{3}}{\left(1 - m\right)^{3}}.
\end{aligned}
\end{equation}

For this cambered airfoil, the thickness needs to be applied perpendicular to the camber line. Therefore, the coordinates of the upper and lower airfoil surfaces become:

\begin{equation} 
\begin{cases}
	x_{U}=x-T(x)\mathrm{sin}(\theta(x)),\\
    x_{L}=x+T(x)\mathrm{sin}(\theta(x)),
\end{cases}
\end{equation}

\begin{equation} 
\begin{cases}
y_{U}=y_{c}+T(x)\mathrm{cos}(\theta(x)),\\
y_{L}=y_{c}-T(x)\mathrm{cos}(\theta(x)),
\end{cases}
\end{equation}
where \(\theta\) is obtained from derivative of \(y_{c}\)
\begin{equation}
\begin{aligned} \label{theta}
	\theta(x) = \mathrm{arctan}(\frac{dy_{c}}{dx}).
\end{aligned}
\end{equation}

\medskip

\bibliographystyle{elsarticle-num}
\bibliography{mybibliography}

\end{document}